\documentclass[iop]{emulateapj}

\usepackage{epsfig, longtable, natbib}
\newcommand{\kms}{km s$^{-1}$}
\newcommand{\ie}{i.e.}
\newcommand{\eg}{e.g.}
\newcommand{\rmin}{$\mathcal{R}_{min}$}
\newcommand{\sigmav}{\sigma_\mathrm{v}}
\newcommand{\siggal}{\sigma_\mathrm{v}^\mathrm{gal}}
\newcommand{\sigdm}{\sigma_\mathrm{v}^\mathrm{DM}}
\newcommand{\Nobs}{N_\mathrm{obs}}

\begin{document}



\title{The DEEP2 Galaxy Redshift Survey: \\
The Voronoi-Delaunay Method catalog of galaxy groups}


\author{Brian F. Gerke\altaffilmark{1,2}, Jeffrey
  A. Newman\altaffilmark{3}, Marc Davis\altaffilmark{4}, Alison L.
  Coil\altaffilmark{5}, Michael C. Cooper\altaffilmark{6}, Aaron
  A. Dutton\altaffilmark{7}, S.~M.
  Faber\altaffilmark{8}, Puragra Guhathakurta\altaffilmark{8}, Nicholas
  Konidaris\altaffilmark{9}, David C. Koo\altaffilmark{8}, Lihwai
  Lin\altaffilmark{10}, Kai Noeske\altaffilmark{11}, Andrew
  C. Phillips\altaffilmark{8}, David J. Rosario\altaffilmark{12}, Benjamin
  J. Weiner\altaffilmark{13}, Christopher
  N. A. Willmer\altaffilmark{13}, Renbin Yan\altaffilmark{14}}

\altaffiltext{1}{KIPAC, SLAC National Accelerator Laboratory, 2575 Sand Hill
  Rd. MS 29, Menlo Park, CA  94725}
\altaffiltext{2}{Present address: Lawrence Berkeley National Laboratory, 1 Cyclotron Rd. MS 90-4000, Berkeley, CA 94720}
\altaffiltext{3}{Department of Physics and Astronomy, 3941 O'Hara St.,
  Pittsburgh, PA 15260}
\altaffiltext{4}{Department of Physics and Department of Astronomy, Campbell Hall, University of
  California--Berkeley, Berkeley, CA 94720}
\altaffiltext{5}{Center for Astrophysics and Space Sciences,
  University of California, San Diego, 9500 Gilman Dr., MC 0424, La
  Jolla, CA 92093}
\altaffiltext{6}{Center for Galaxy Evolution, Department of Physics and
  Astronomy, University of California--Irvine, Irvine, CA 92697}
\altaffiltext{7}{Department of Physics and Astronomy, University of
  Victoria, Victoria, BC, V8P 5C2, Canada} 
\altaffiltext{8}{UCO/Lick Observatory, University of California--Santa
  Cruz, Santa Cruz, CA 95064}
\altaffiltext{9}{Caltech 249-17; Pasadena, CA 91125}
\altaffiltext{10}{Institute of Astronomy \& Astrophysics, Academia Sinica, Taipei 106, Taiwan}
\altaffiltext{11}{Space Telescope Science Institute, 3700 San Martin Dr., Baltimore, MD 21218}
\altaffiltext{12}{Max Planck Institute for Extraterrestrial Physics,
  Giessenbachstr. 1, 85748 Garching bei M\"{u}nchen, Germany}
\altaffiltext{13}{Steward Observatory, University of Arizona, 933 N Cherry Ave
Tucson, AZ 85721}
\altaffiltext{14}{Department of Astronomy and Astrophysics, University
  of Toronto, 50 St. George Street, Toronto, ON, M5S 3H4, Canada}

\begin{abstract}
  We present a public catalog of galaxy groups constructed from the
  spectroscopic sample of galaxies in the fourth data release from the
  DEEP2 Galaxy Redshift Survey, including the Extended Groth Strip
  (EGS).  The catalog contains 1165 groups with two or more members in
  the EGS over the redshift range $0<z<1.5$ and 1295 groups at $z>0.6$
  in the rest of DEEP2. $25\%$ of
  EGS galaxies and $14\%$ of high-z DEEP2 galaxies are assigned to
  galaxy groups.  The groups were detected using the Voronoi-Delaunay
  Method, after it has been optimized on mock DEEP2 catalogs following
  similar methods to those employed in ~\citet{Gerke05}.  In the
  optimization effort, we have taken particular care to ensure that
  the mock catalogs resemble the data as closely as possible, and we
  have fine-tuned our methods separately on mocks constructed for the
  EGS and the rest of DEEP2.  We have also probed the effect of the
  assumed cosmology on our inferred group-finding efficiency by
  performing our optimization on three different mock catalogs with
  different background cosmologies, finding large differences in the
  group-finding success we can achieve for these different mocks.
  Using the mock catalog whose background cosmology is most consistent
  with current data, we estimate that the DEEP2 group catalog is 72\%
  complete and 61\% pure (74\% and 67\% for the EGS) and that the
  group-finder correctly classifies 70\% of galaxies that truly belong
  to groups, with an additional 46\% of interloper galaxies
  contaminating the catalog (66\% and 43\% for the EGS).  We also
  confirm that the VDM catalog reconstructs the abundance of galaxy
  groups with velocity dispersions above $\sim 300$\kms, to an
  accuracy better than the sample variance, and that this successful
  reconstruction is not strongly dependent on cosmology.  This makes
  the DEEP2 group catalog a promising probe of the growth of cosmic
  structure that can potentially be used for cosmological tests.
\end{abstract}


\section{Introduction}
\label{sec:intro}

The spherical or ellipsoidal gravitational collapse of an overdense
region of space in an expanding background is a simple dynamical
problem that can be used as an \emph{Ansatz} to predict the mass
distribution of massive collapsed structures in the Cold Dark
Matter cosmological paradigm, as a function of the cosmological
parameters \citep{PS74, BBKS, ST02}. This has led to the widespread
use galaxy clusters and groups as convenient cosmological 
probes.  In addition, it has long been apparent that the galaxy
population in groups and clusters differs in its properties from the
general population of galaxies \citep[\eg, ][]{SB51, Dressler80} and
that the two populations exhibit different evolution \citep{BO84}.
This suggests that galaxy groups and clusters can be used as
laboratories for studying evolutionary processes in galaxies.  For
both of these reasons, a catalog of groups and clusters has been
derived for every large survey of galaxies.

The history of  group and cluster finding in galaxy surveys includes
a wide variety of detection methods, starting with the visual
detection of local clusters in imaging data by \citet{Abell58}.
The approaches can be broadly divided into two categories: those that
use photometric data only, and those that use spectroscopic redshift
information.  In relatively shallow photometric data, it is possible
to find clusters by simply looking for overdensities in the on-sky
galaxy distribution, but in modern, deep photometric surveys,
foreground and background objects quickly overwhelm these
density peaks at all but the lowest redshifts.  Recent photometric
cluster-finding algorithms thus typically also rely on assumptions
about the properties of galaxies in clusters, on photometric redshift
estimates, or on a combination of the two \citep[\eg, ][]{Postman96,
  GY00, Koester07a, Li08, Liu08, Adami10,
  Hao10, Milkeraitis10, Soares-Santos11}.  

Spectroscopic galaxy redshift surveys remove much of the problem of
projection effects from cluster-finding efforts, though not all of
it, owing to the well-known finger-of-god effect.  Since closely
neighboring galaxies in redshift space can be assumed to be physically
associated, it is possible to use spectroscopic surveys to reliably
detect relatively low-mass, galaxy-poor systems (\ie, galaxy groups),
in addition to rich, massive clusters.  The most popular approach
historically has been the
friends-of-friends, or percolation, algorithm, which links galaxies
together with their neighbors that lie within a given linking length
on the sky and in redshift space, without reference to galaxy
properties.  This technique was pioneered in the CfA redshift survey
\citep{HG82} and is still in common use in present-day redshift
surveys \citep[\eg, ][]{Eke04, Berlind06, Knobel09}.  Recently, other
redshift-space algorithms have also had success by including simple
assumptions about the properties of galaxies in clusters and groups
\citep[\eg, ][]{Miller05, Yang05}.  The primary disadvantage of
cluster-finding in redshift-space data is that spectroscopic surveys
generally cannot schedule every galaxy for observation, leading to a
sparser sampling of the galaxy population than is available in
photometric data.  When the sampling rate becomes extremely low,
standard methods like friends-of-friends have a very high failure
rate.  This is a particular concern for high-redshift surveys, for
which spectroscopy is very observationally expensive.

In any case, since cluster-finding algorithms search
for spatial associations in a pointlike dataset, it can be shown that
a perfect reconstruction of the true, underlying bound systems can
never be achieved owing to random noise \citep{SzSz96}.  Indeed, it
has long been known that a fundamental trade-off exists between the
purity and completeness of a cluster catalog when compared with the
underlying dark-matter halo population in N-body models \citep{NW87}:
a catalog cannot be constructed that detects all existing clusters and
is free of false detections.  In order to fully understand and
minimize these
inevitable errors, it has become standard practice to
use mock galaxy catalogs, based on $N$-body dark-matter simulations, to
test cluster-finding algorithms, optimize their free parameters, and
estimate the level of error in the final catalog.  In all such
studies, some effort has been made to ensure that the mock
catalogs resemble the data at least in a qualitative sense, but little
work has been done to examine how quantitative differences between the
mocks and the data, or inaccuracies in the assumed background
cosmology, will impact the group-finder calibration.

In this paper, we present a catalog of galaxy groups and clusters for
the final data release (DR4) of the DEEP2 Galaxy Redshift Survey
\citep{DEEP2}, a spectroscopic survey of tens of thousands of mostly
high-redshift galaxies, with a median redshift around $z=0.9$.  The
catalog is made available to the public on the DEEP2 DR4
webpage\footnote{\tt http://deep.berkeley.edu/dr4}.  To construct this
catalog, we make use of the the Voronoi-Delaunay Method (VDM)
group-finder, which was originally developed by \citet{MDNC02} for use
in relatively sparsely sampled, high-redshift surveys similar to
DEEP2. To test and calibrate our methods, we make use of a set of
realistic mock galaxy catalogs that we have recently constructed for
DEEP2 (Gerke et al. in preparation).  These catalogs have been
constructed for several different background cosmologies, allowing us
to test the impact of cosmology on the group-finder calibration and
error rate.  This work updates and expands upon the group-finding
efforts of \citet{Gerke05} (hereafter G05), who detected groups with
the VDM algorithm in early DEEP2 data using an earlier set of DEEP2
mocks for calibration.

Our goals in constructing this catalog are similar to the historical
ones described above.  First, a catalog of galaxy groups is an
interesting tool for studying the evolution of the galaxy population
in DEEP2, as well as for studying the baryonic astrophysics of groups
and clusters themselves, as has been demonstrated in various papers
using the G05 catalog \citep{Fang06, Coil06a, Gerke07a, Georgakakis08,
  Jeltema09}. In addition, it has been shown that a catalog of groups
from a survey like DEEP2 can be used to probe cosmological parameters,
including the equation of state of the dark energy, by counting groups
as a function of their redshift and velocity dispersion
\citep{NMCD02}; we aim to produce a group catalog suitable for that
purpose here.

We proceed as follows.  In Section~\ref{sec:data} we introduce the
DEEP2 dataset and describe our methods for constructing realistic
DEEP2 mock catalogs with which to test and refine our group-finding
methods.  Section~\ref{sec:methods} details the specific criteria we
use for such testing.  In Section~\ref{sec:VDM} we give a brief
overview of VDM, including some changes to the G05 algorithm, and we
optimize the algorithm on our mock catalogs in
Section~\ref{sec:optimize}.  The latter section also explores the
dependence of our optimum group-finding parameters on the assumed
cosmology of the mock catalogs.  Section~\ref{sec:catalog} presents
the DEEP2 group catalog and compares it to other high-redshift
spectroscopic group catalogs.  Throughout this paper, where necessary
and not otherwise specified, we assume a flat $\Lambda$CDM cosmology
with $\Omega_M = 0.3$ and $h=0.7$.

\section{The DEEP2 survey and  mock catalogs}
\label{sec:data}
\subsection{The DEEP2 dataset}
\label{sec:DEEP2}
The DEEP2 (Deep Extragalactic Evolutionary Probe 2) Galaxy Redshift
Survey is the largest spectroscopic survey of homogeneously selected
galaxies at redshifts near unity.  It consists of some 50,000 spectra
obtained in one-hour exposures with the DEIMOS spectrograph
\citep{Faber03} on the Keck II telescope.  This dataset yielded more
than 35,000 confirmed galaxy redshifts; the rest were either stellar
spectra or failed to yield a reliable redshift identification.  DEEP2
will be comprehensively described in \citet{DEEP2}; most details of
the survey can also be found in \citet{Willmer06}, \citet{DGN04}, and
\citet{AEGIS}.
Here we summarize the main survey characteristics, focusing on issues
of particular importance for group finding.

DEEP2 comprises four separate observing fields, chosen to lie in
regions of low Galactic dust extinction that are also widely separated
in RA to allow for year-round observing.  With a combined area of
approximately three square degrees, the DEEP2 fields probe a volume of
$5.6\times 10^6\,h^{-3}$ Mpc$^3$ over the primary DEEP2 redshift range
$0.75<z<1.4$.  This is an excellent survey volume for studying galaxy
groups: at the relevant epochs, one expects to find more than one
thousand dark matter halos with masses in the range of galaxy groups
(roughly $5\times 10^{12} M_\odot \la M_{halo} \la 1\times 10^{14}
M_\odot$) in a volume of this size.  DEEP2 is less well suited for
studying clusters: at most there should be a few to a few tens of halos with
cluster masses ($M_{halo}\ga 1\times 10^{14} M_\odot$) in the DEEP2
fields.  Since our final catalog will be dominated by objects that are
traditionally referred to as \emph{groups} (rather than clusters) we
will use that term throughout this work as a shorthand to refer to
both groups and clusters.

DEEP2 spectroscopic observations were carried out using the 1200-line
diffraction grating on DEIMOS, giving a spectral resolution of $R\sim
6000$.  This yields a velocity accuracy of $\sim 30$ km/s (measured
from repeat observations of a subset of targets).  Such high-precision
velocity measurements make DEEP2 an excellent survey for detecting
galaxy groups and clusters in redshift space, which is our strategy
here.  The velocity errors are substantially smaller than typical
galaxy peculiar velocities in groups, so the dominant complication for
redshift-space group-finding will be the finger-of-god effect, rather
than redshift-measurement error.

Targets for DEIMOS spectroscopy were selected down to a limiting
magnitude of $R=24.1$ from three-band ($BRI$) photometric observations
taken with the CFH12k imager on the Canada-France-Hawaii Telescope
\citep{Coil04b}.  To focus the survey on typical galaxies at $z\sim 1$
(rather than low-$z$ dwarfs) most DEEP2 targets were also restrictied to
a region of $B-R$ versus $R-I$ color-color space that was chosen to
contain a nearly complete sample of galaxies at $z>0.75$
\citep{DGN04}.  Tests with spectroscopic samples observed with no
color pre-selection show that the DEEP2 color cuts exclude the bulk of
low-redshift targets, while still including $\sim 97\%$ of galaxies in
the range $0.75<z<1.4$ \citep{DEEP2}.  (At $z>1.4$---in the so-called
``redshift desert''---it is difficult to obtain successful galaxy
redshifts because of a lack of spectral features in the
observed optical waveband.)

Despite the high completeness of the DEEP2 color selection at high
redshift, there remain a number of observational effects that reduce
the sampling density of galaxies in groups and clusters.  The simplest
is the faint apparent magnitude range of $z\sim 1$ galaxies.  DEEP2 is
limited to luminous galaxies ($L\ga L*$; \citealt{Willmer06}) at most
redshifts of interest; even massive clusters will contain a few tens
of such galaxies at most.  At redshifts near $z=0.9$, DEEP2 has a
number density of galaxies $n\sim 0.01$ (Newman et al. in
preparation), corresponding to a fairly sparse galaxy sample with mean
intergalaxy separation $s\sim 5 \mathrm{h}^{-1}\mathrm{Mpc}$ (comoving
units). The DEEP2 group sample will thus be made up of systems with
relatively low richnesses.

A further complication arises
from the effects of k-corrections on high-redshift galaxies, which
translate the DEEP2 $R$-band apparent magnitude limit into a an
evolving, color-dependent luminosity cut in the rest frame of DEEP2
galaxies.  As discussed in detail in \citet{Willmer06} and
\citet{Gerke07a}, red-sequence galaxies in DEEP2 will have a brighter
absolute magnitude limit than blue galaxies at the same redshift, and
this disparity increases rapidly with redshift as the observed $R$
band shifts through the rest-frame $B$ band and into the $U$ band
(\emph{cf.}  Figure 2 of \citealt{Gerke07a}).  Galaxies on the red
sequence are well known to preferentially inhabit the overdense
environments of groups and clusters, and this relation holds at $z\sim
1$ in DEEP2 \citep{Cooper07, Gerke07a}. This means that groups and
clusters of galaxies in DEEP2 will have a lower sampling density than
the overall galaxy population, and the observed galaxy population in
groups will be skewed toward more luminous objects.

Further undersampling of DEEP2 group galaxies results from the
unavoidable realities of multiplexed spectroscopy.  
DEEP2 spectroscopic targets were observed using custom-designed DEIMOS
slitmasks that allowed for simultaneous observations of more than $100$
targets.  Although slits on DEEP2 masks could be made
as short as $3^{\prime\prime}$, and some slits could be designed to
observe two neighboring galaxies at once, the requirement that slits
not overlap with one another along the spectral direction of a mask
inevitably limits the on-sky density of targets that can be observed.
Overall, DEEP2 observed roughly $65\%$  of
potential targets, but this fraction is necessarily lower in crowded
regions on the sky owing to slit conflicts.  The adaptive DEEP2
slitmask-tiling strategy relieves crowding issues somewhat, since each
target has two chances for selection on overlapping slitmasks, but
there is still a distinct anticorrelation between targeting rate and
target density: the sampling rate for targets
in the most crowded regions on the sky is roughly $70\%$ of the median
sampling rate  (G05, \citealt{DEEP2}).

\begin{figure}
\epsfig{width=3.25in, file=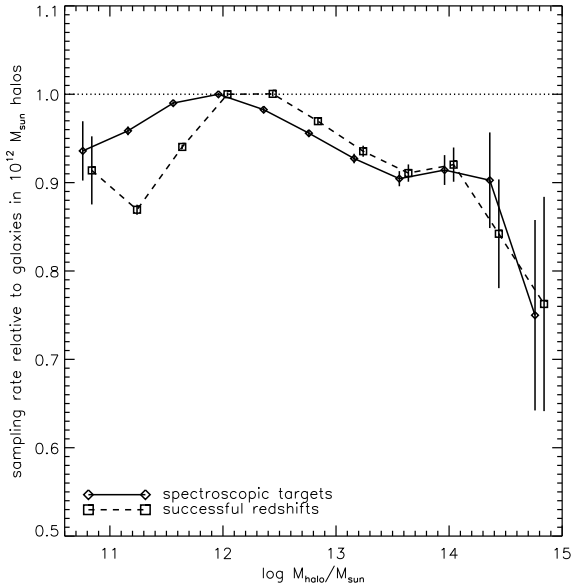}
\caption{Relative spectroscopic targeting and redshift-success rates
  of mock DEEP2 galaxies versus parent halo mass.  Because of
  increased slit conflicts in crowded regions, galaxies in groups and
  clusters ($M_{halo}>10^{13} M_\odot$) are targeted for DEIMOS
  spectroscopy at a lower rate than galaxies like the Milky Way
  ($M_{halo}\sim 10^{12}M_\odot$) (solid line).  The effect is mild,
  however, and remains less than $20\%$ for all but the most massive
  clusters in the mock catalog.  The sampling rate also falls for
  low-mass halos, since these contain faint galaxies, some fraction of
  which fall below the DEEP2 magnitude limit.  Faint red galaxies are
  preferentially undersampled further because they are less likely to
  yield reliable redshifts; however, this effect is also mild and in
  any caes is
  limited mostly to galaxies in low-mass halos (dashed line).}
\label{fig:sampling}
\end{figure}

Nevertheless, as discussed in G05, the significant line-of-sight
distance covered by DEEP2 means that high-density regions \emph{on the
  sky} do not necessarily correspond to high-density regions in
three-space.  The impact of slit conflicts on the sampling of groups
and clusters should therefore be lower than the effect seen in crowded
regions of sky.  We can test this explicitly using simulated galaxies
in the mock catalogs described in Section~\ref{sec:mocks}.  Galaxies
in the mocks are selected using the same slitmask-making algorithm
as used for DEEP2, and since the mocks also contain
information on dark-matter halo masses, it is possible to investigate
the effect of this algorithm on the sampling rate in group-mass and
cluster-mass halos.  As shown in Figure~\ref{fig:sampling}, galaxies
in massive halos are undersampled relative to field galaxies, but the
effect is modest, amounting to less than a $10\%$ reduction in
sampling rate at group masses and only a $\sim 20$\% reduction for the
most massive clusters in the mocks.

Redshift failure is a final factor that impacts the sampling rate of
groups and clusters.  After visual inspection, roughly $30\%$ of DEEP2
spectra fail to yield a reliable redshift (i.e., do not receive DEEP2
redshift quality flag 3 or 4, which correspond to 95\% and 99\%
confidence in the redshift identification, respectively). These
redshift failures are excluded from all samples used for group-finding.
Follow-up observations (C. Steidel, private communication) show that
roughly half of these redshift failues lie at $z>1.4$, but the
remainder serve to further reduce the DEEP2 sampling rate in the
target redshift range.  The redshift failure rate increases sharply
for galaxies in the faintest half magnitude of the sample, and it is
also boosted for red galaxies, since these tend to lack strong
emission lines, making redshift identification more difficult.  One
might expect that this would decrease the sampling rate preferentially
in groups and clusters, which should have a large number of faint red
satellite galaxies.  It is also possible to test this with the mock
catalogs (which account for the color and magnitude-dependence of the
incompleteness, as discussed in the next section).  As shown in
Figure~\ref{fig:sampling} (dashed curve), redshift failures have a
stronger effect on mock galaxies in low-mass halos (since they
preferentially host faint galaxies) than in groups and clusters so
that, if anything, the relative sampling rate of groups and clusters
is boosted slightly by redshift failures.

In any case, Figure~\ref{fig:sampling} demonstrates the
importance of having realistic mock catalogs on which to calibrate
group-finding methods.  Without accurate modeling of the selection
probability for galaxies in massive halos relative to field galaxies,
it will be difficult to have confidence in measures of group-finding
success (\eg, the completeness and purity of the group catalog).  In the
next section, we will describe the mock catalogs we use to test our
group-finding methods and optimize them for the DEEP2 catalog,
focusing on the steps that have been taken to account for all of the
different DEEP2 selection effects discussed above.

\subsubsection{The Extended Groth Strip}

Before we proceed, it is important to describe the somewhat different
selection criteria that were used in one particular DEEP2 field, the
Extended Groth Strip (EGS).  This field is also the site of AEGIS, a
large compendium of datasets spanning a broad range of wavelengths,
from X-ray to radio \citep{AEGIS}.  To maximize the redshift coverage
of these multiwavelength datasets, DEEP2 targets were selected without
color cuts, so that galaxy spectra are obtained across the full
redshift range $0<z<1.4$.  However, spectroscopic target selection
used a probabilistic weighting as a function of color, to ensure
roughly equal numbers of targets at low and high redshift; this means
that the sampling rate of galaxies will vary differently with redshift
than would be expected in a simple magnitude-limited sample.
Furthermore, the EGS was observed with a different spectroscopic
targeting strategy, so that each galaxy has four chances to be
observed on different overlapping slitmasks.  The overall sampling
rate in the EGS is thus boosted somewhat relative to the rest of
DEEP2.  These differences in selection mean that it will be important
to calibrate our group-finding techniques separately for the EGS and
the rest of the DEEP2 sample.  Our mock catalogs will therefore need
to be flexible enough to account for the differences in selection
between the EGS and the rest of DEEP2.

\subsection{DEEP2 mock catalogs}
\label{sec:mocks}

The success of any group-finder will depend sensitively on the
selection function of galaxies in halos of different masses, since
this drives the observed overdensity of groups and clusters relative
to the background of field galaxies.  It is therefore crucial to test
and optimize group-finding algorithms on simulated galaxy catalogs
that capture and characterize this mass-dependent selection function
as accurately as possible.  It is helpful to couch this discussion in
the terminology of the halo model \citep[\eg, ][]{PS00, Seljak00,
  Ma00}, particularly the halo occupation distribution (HOD)
$\bar{N}(M)$, which is the average number of galaxies meeting some
criterion (usually a luminosity threshold) in a halo of mass $M$.
What we would like is a mock catalog that correctly reproduces the HOD
of \emph{observed} galaxies, not just the HOD for galaxies above some
luminosity cut, in group-mass halos.  As discussed below, this will
require us to improve upon the mocks we used for the initial DEEP2
group-finding calibration in G05.

In that study, we optimized the VDM group-finder using the mock
catalogs of \citet{YWC04} (hereafter YWC). Those authors produced mock
DEEP2 catalogs from a large-volume N-body simulation by adding
galaxies to dark-matter halos according to a conditional luminosity
function $\Phi(L|M)$ whose form and parameters were chosen to be
consistent with the \citet{Coil04} galaxy autocorrelation function
measured in early DEEP2 data.  Since the HOD is directly linked to the
correlation function, this implied that the HOD in the
mocks was consistent with existing data.  However, the agreement
between the high-redshift mock and measured correlation functions was
marginal at best, and later measurements \citep{Coil06b} narrowed the
error bars on the DEEP2 correlation function so that the existing
mocks no longer agree with the data at high redshift.  Indeed, direct
modeling of the HOD from the DEEP2 correlation function \citep{ZCZ07}
is quite inconsistent with the HOD that was used in YWC.  In
particular, the YWC HOD had a power-law index of $\sim 0.7$ at high
masses, while the HOD derived from DEEP2 data has a power-law index
near unity.  This suggests that the galaxy occupation of groups in the
YWC mocks is quite different from that in the real universe.  

Another difficulty arises when we consider color-dependent selection
effects.  As discussed above, the DEEP2 magnitude limit translates
into an evolving, color-dependent luminosity cut that also evolves
with redshift, which may lead to preferential undersampling of groups
and clusters.  This is further complicated by the fact that the
color-density relation also evolves over the DEEP2 redshift range
\citep{Gerke07a, Cooper07}.  Correct modeling of galaxy colors in the
mock catalogs is therefore critical to proper calibration of our
cluster-finding efforts.  Unfortunately, the YWC mocks did not contain
any color information, so any preferential color-dependent
undersampling of groups and clusters was not reflected there.
\citet{Gerke07a} addressed this problem by adding colors to the YWC
mocks according to the measured DEEP2 color-density relation from
\citet{Cooper06a}, but this did not address the inaccuracy of the
underlying HOD.

A final possible problem involves the choice of cosmological
background model used to construct the mock catalogs.  The YWC mocks
we used in G05 used N-body simulations calculated in a flat,
$\Lambda$CDM cosmology with parameters $\Omega_M = 0.3$ and $\sigma_8
= 0.9$, both of which lie outside the region of parameter space
preferred by current data.  Because changing these parameters has a
significant impact on the halo abundance at $z\sim 1$, and because any
realistic mock catalog will be constrained to match the abundance of
galaxies, changes in the cosmology will necessarily have a substantial
impact on the HOD.  For example, a model with a higher (lower)
$\sigma_8$ will have a higher (lower) abundance of halos at any given
mass, and thus will require a lower (higher) $\bar{N}(M)$ to match the
observed galaxy abundance.  This effect will be discussed in more
depth in the paper describing the new DEEP2 mocks (Gerke et al. in
prep.), but here it will be important to assess its impact on group
finding.

Thus, as pointed out in YWC, it is important to update the mock
catalogs to match DEEP2 more closely, now that a larger dataset
is available.  In this paper we make use of a new set of DEEP2 mock
catalogs that remedy many of the inadequacies of the previous mocks.
These mocks will be described in detail in a paper by Gerke et al. (in
preparation); here we summarize the most important improvements over
YWC for the purposes of group-finding calibration.

The new mocks are produced from N-body simulations that have
sufficient mass resolution to detect dark-matter halos and subhalos
down to the mass range of dwarf galaxies with absolute magnitudes
$\sim M^\ast+10$.  This permits us to assign galaxies uniquely to
dark-matter halos and subhalos over the full range of redshift and
luminosity covered by DEEP2, including the EGS.  In order to
investigate the impact of different cosmological models on group
finding, we have constructed mock catalogs using three different
simulations with three different background cosmologies that span the
current range of allowed models; these are summarized in
Table~\ref{tab:mockcosmo}.  We use the mocks constructed from the
Bolshoi simulation \citep{Bolshoi} as our fiducial model for quoting
our main results, since its parameters are most consistent with
current data, but we will use the other two cosmological models to
investigate the impact on our results of changes in the cosmological
background.  As discussed in Gerke et al. (in prep.), we construct
light cones from these simulations, each having the geometry of a
single DEEP2 observational field.  To properly account for cosmic
evolution, we stack different simulation timesteps along the line of
sight, and we limit the number of lightcones we create for each
simulation to ensure that the resulting mocks sample roughly
independent volumes at fixed redshift.

\begin{deluxetable}{lccccc}
\tabletypesize{\small}
\tablewidth{0pt}
\tablecaption{Summary of the simulations used to construct DEEP2 mock catalogs.}
\tablehead{
\colhead{Simulation}&\colhead{Box size}\tablenotemark{a}&\colhead{Fields}\tablenotemark{b}&\colhead{$\Omega_M$}&\colhead{$\sigma_8$}& $h$
}
\startdata
Bolshoi  & 250 & 40 &0.27 & 0.82 & 0.7 \\
L160 ART  & 160 & 12 & 0.24   & 0.7   & 0.7 \\
L120 ART   & 120 & 12 & 0.3   & 0.9   & 0.73 \\
\enddata
\tablenotetext{a}{comoving $h^{-1}$ Mpc on a side.}
\tablenotetext{b}{Number of mock 1 deg$^2$ DEEP2 fields or 0.5 deg$^2$
  fields produced from each simulation.}
\label{tab:mockcosmo}
\end{deluxetable}

To add mock galaxies to these dark-matter-only lightcones, we use the
so-called subhalo abundance-matching approach
\citep[\eg,][]{Conroy06a, VO06} to assign galaxy luminosities to
dark-matter subhalos identified in the simulations.  Using the
measured DEEP2 galaxy luminosity function (including its redshift
evolution) and simulated subhalo internal velocity-dispersion
function, we map galaxy luminosities to subhalos at fixed number
density.  By contrast, the dark-matter simulations used for the YWC
mocks did not include detections of dark-matter substructures to a
sufficiently low mass, so galaxies were assigned to dark-matter halos
stochastically from an HOD, with satellite galaxies assigned to
randomly selected dark matter particles. Our subhalo-based procedure
should give a more accurate representation of the luminous profiles
and galaxy kinematics of galaxy clusters than the \citet{YWC04} mocks.
In addtion, the simulations used for the earlier mocks resolved halo
masses sufficient to host central galaxies only down to $\sim 0.1
L^\ast$.  This made it impossible to create realistic mock catalogs
for the EGS field, since this region includes faint dwarf galaxies at
low redshits.  Our new mocks resolve halos and subhalos to masses low
enough to accommodate all DEEP2 galaxies except for a handful of very
faint dwarfs at $z\la 0.05$.

\citet{Conroy06a} showed that the abundance-matching procedure
reproduces the galaxy autocorrelation function at a wide range of
redshifts, provided that the subhalo velocity function uses the
subhalo velocities as measured at the moment they were accreted into
larger halos. and for a particular choice of cosmological parameters
that is now disfavored by the data.  As discussed in Gerke et al (in
prep.), however, for the more accurate cosmology used in Bolshoi, the
abundance-matching approach does not reproduce the DEEP2 projected
two-point function at $z\sim 1$, lying some $20$--$40\%$ higher than
the measurement from \citep{Coil06b}.  As we also discuss in that paper,
the likely resolution to this discrepancy would involve an
abundance-matching appoach that includes scatter in luminosity at
fixed subhalo velocity dispersion, with larger scatter at lower
dispersion values.  This is likely to mainly impact the HOD at low
masses, near the transition of $\bar{N}(M)$ between zero and unity,
while causing minimal alteration in the HOD at group and cluster
masses.  Since the Bolshoi mock HOD matches the measured \citet{ZCZ07}
HOD well at these masses,  we concluded that the clustering  mismatch
does not preclude using these mocks for group-finder optimization. The
overall occupation of group-mass halos in the mocks should represent
the real universe well.  What then remains is to account for the
various observational selection effects that translate this into an
observed HOD for groups.

To add galaxy colors to the mocks, we have followed an approach
similar to the one used in \cite{Gerke07a} (which was itself inspired
by the ADDGALS algorithm; Wechsler et al. in prep.).  We assign a
rest-frame $U-B$ color to each mock galaxy by drawing a DEEP2 galaxy
with similar redshift, luminosity, and local galaxy overdensity. While
performing the color assignment, we must also account for galaxies
that fall below the DEEP2 apparent magnitude limit.  At fixed redshift
redshift, there is some luminosity range in which the DEEP2 sample is
partially incomplete, depending on galaxy color.  In these luminosity
ranges, we select galaxies for exclusion from the mock catalog
depending on their local density, until the local density distribution
in the mock is consistent with the measured distribution in DEEP2.
This technique effectively uses local galaxy density as a proxy for
color and ensures that the impact of the DEEP2 selection function on
the sampling of galaxy environment is accurately reproduced in the
mocks.  Full details of the color-assignment algorithm (which are
somewhat complex and beyond the scope of this discussion) can be found
in the paper describing the mock catalogs (Gerke et al. in
preparation).

After assigning rest-frame colors, we then assign observed apparent
$R$-band magnitudes by inverting the $k$-correction procedure of
\citet{Willmer06}; this procedure accurately reproduces the evolving,
color-dependent luminosity cut that is imposed by the DEEP2 magnitude
limit, as well as the color-density relation, so any undersampling of
groups and clusters owing to color-dependent selection effects should
also be captured in these mocks.

As we did in G05, to simulate the effects of DEEP2 spectroscopic
target selection we pass our mock catalogs through the same
slitmask-making algorithm that was used to schedule objects for DEEP2
observations \citep{DGN04, DEEP2}.  The DEEP2 color cuts do not
give a completely pure sample of high-redshift galaxies, so the pool
of mock targets for maskmaking also includes foreground ($z<0.75$) and
background ($z>1.4$) galaxies, as well as randomly positioned stars,
in proportions that are consistent with those found in the DEEP2 sample.
To make mocks of the EGS field, we use the somewhat different
target-selection algorithm that was used for the EGS, including
galaxies at all redshifts, but giving higher selection probability to
galaxies at $z>0.75$ in a manner that reflects the color-dependent
weighting applied to the real EGS.  Any density-dependent effects on
the sampling rate that are driven by slit conflicts should therefore
be fully accounted for in the mocks.

As a final step,
we must replicate the effects of DEEP2 redshift failures, as a
function of galaxy color and magnitude.  To do this, we utilize the
incompleteness-correction weighting scheme devised by
\cite{Willmer06}.  This scheme assigns a weight to each galaxy
according to the fraction of similar galaxies (in observed
color-color-magnitude space) that failed to yield a redshift.  When we
add colors to the mock galaxies by selecting galaxies from the DEEP2
sample, we also assign each mock galaxy the incompleteness weight
$w_i$ of the DEEP2 galaxy we have drawn (with some small corrections,
described in Gerke et al. in preparaion).
Although this was intended to correct for redshift incompleteness in
the data, it can be inverted to \emph{produce} incompleteness in the mock:
after we have selected targets with the DEEP2 slitmask-making 
algorithm, we reject $\sim 30\%$ of these targets, with a rejection
probability given by $1/w_i$.  This procedure naturally reproduces
any dependence of the DEEP2 redshift-success rate on galaxy color and
magnitude.

These mock catalogs accurately reproduce a wide range of statistical
properties of the DEEP2 dataset (Gerke et al. in prep.).  Most
importantly for group-finding efforts, though the mocks match
\emph{(1)} The HOD at group masses ($M\ga 5\times 10^{12}$), as
measured in \citet{ZCZ07} for several different luminosity
thresholds,\emph{(2)} the evolving color-density relation that was
measured in \citet{Cooper06a} and \citet{Cooper07}, and \emph{(3)} the
redshift distribution of the DEEP2 data. These three points of
agreement should be sufficient to ensure that the \emph{observed}
DEEP2 HOD for group-mass halos is accurately reproduced by the mocks.
We can thus proceed with confidence in using these mocks to optimize
our group-finding techniques.

\subsubsection{The effects of DEEP2 selection on the observed group population}
\begin{figure*}
\centering
\epsfig{width=7in, file=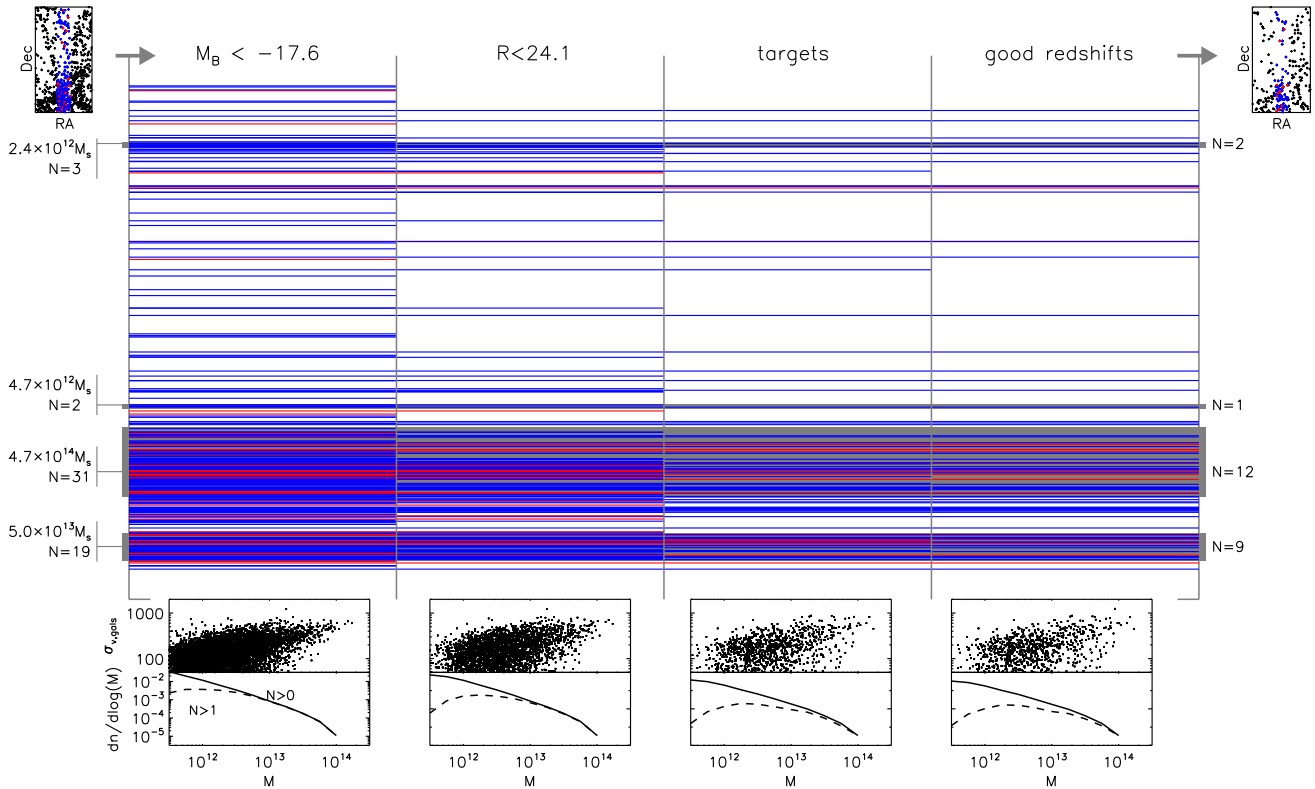}
\caption{ The dilution of a small subregion of a DEEP2 mock catalog by
  observational selection effects.  Galaxies have been selected from a
  small subregion of a DEEP2 mock catalog, roughly 4 arcmin in RA by
  30 arcmin in Dec with a redshift depth of 0.05.  This strip is
  indicated in projection on the sky by the colored points in the
  panels at top left (before selection) and after selection at top
  right (afterward).  Galaxies on the red sequence (\ie redder than
  the red-blue divide given in \citet{Willmer06}) are indicated in
  red, and blue galaxies are shown in blue.  In order to show the
  impact on cluster selection, this narrow slice in RA, Dec and
  redshift was chosen to contain the most massive high-redshift
  cluster in the mock catalogs (a $4.7\times 10^{14} M_\odot$ object
  at $z=0.8$).  The main panel is a schematic diagram of these
  galaxies' path through the DEEP2 selection process.  At left, we
  begin with horizontal lines (arranged vertically in order of the
  declination coordinate) representing all galaxies in this subregion
  more luminous than $M_B = -17.6$.  Each vertical grey line
  represents a step in the DEEP2 selection procedure; galaxies are
  excluded from the sample by the $R=24.1$ apparent magnitude limit,
  by the targeting procedure for assigning galaxies to DEIMOS slits,
  and by failures to obtain good redshifts for some observed galaxies.
  Horizontal gray bands in the main panel indicate the spatial extent
  of the four most massive halos in this small region (note that the
  colored lines within these gray bands are not necessarily all
  members of these halos, owing to projection effects).  The masses of
  the halos are indicated, as are their richnesses before and after
  dilution by DEEP2 selection processes.  The bottom panels show the
  aggregate impact of galaxy selection effects on the halo and group
  population.  The lower half of each panel shows the mass function of
  halos containing one or more galaxy in each sample (solid lines) and
  the mass function of groups with two or more members (dashed lines).
  The group population selected in DEEP2 spans a very broad range in
  mass and represents an incomplete halo sample at all but the very
  highest masses.  The upper half of each panel shows the relation
  between halo mass and measured group velocity dispersion $\siggal$
  for all groups with two or more members.  The mean relation remains
  approximately constant, although the scatter increases since there
  are fewer galaxies sampling the velocity field.}
\label{fig:dilution}
\end{figure*}

First, though, it will be interesting to use the mocks to investigate
the impact of observational effects on the galaxy population of
massive halos in DEEP2.  (We also explored this in some detail in G05;
see Figures 2 and 3 of that paper). Figure~\ref{fig:dilution}
summarizes the impact of the various DEEP2 selection effects on
galaxies in massive dark matter halos in a narrow slice through a mock
catalog, which contains the most massive high-redshift halo in the
mocks (this region is depicted in projection on the sky, before and
after selection, as the colored points in the upper left and right
panels, respectively).  There are three primary selection effects that
remove galaxies from the mock sample.  In the figure, these selections
are depicted visually by vertical lines across the main panel, and
galaxies' paths through the selection process are shown by horizontal
lines running from left to right, with group-mass halos indicated by
gray horizontal bands.  First, the DEEP2 $R=24.1$ magnitude
limit removes faint galaxies, with red galaxies being excluded at
brighter luminosities than blue ones. DEIMOS target selection then
removes a random subsample of the remaining galaxies, with some
preferential rejection occurring in massive halos.  Finally, some
galaxies fail to yield redshifts, further diluting the sample.  The
impact of this dilution on the population of galaxies in groups can be
quite strong: the most massive halo shown in the main panel loses some
$60\%$ of its members.  It also introduces an added degree of
stochasticity into the mass-selection of halos.  The least-massive
halo shown in the figure contains two observed galaxies, and would be
identified as a group, while the next most massive halo contains only
one observed galaxy, so it would be identified as an isolated galaxy.

The lower panels in Figure~\ref{fig:dilution} show the effect on the
mass functions of observed galaxies and groups.  DEEP2 selection
effects mean that the sample of systems with two or more observed
galaxies will only be a complete sample of massive halos at relatively
high masses $\ga 5\times 10^{13}M_\odot$.  However, the cutoff in the
mass selection function for groups is quite broad, owing to the
stochastic effects mentioned above, so that even halos with $M<
10^{12} M_\odot$ have some chance of being identified as groups.

The lower panels also show the effect of DEEP2 selection on the
relation between halo mass and observed group velocity dispersion (for
systems with two or more galaxies at each stage).  As expected, the
scatter in this relation increases as we move through the selection
process, since the number of galaxies sampling the velocity field is
reduced.  However, a clear correlation remains between the mass of a
halo and the dispersion $\siggal$ of its galaxies'
peculiar velocities.  It should therefore be possible, at least in
principle, to use a DEEP2 group catalog to measure the halo mass
function and constrain cosmological parameters, as proposed in
\citet{NMCD02}, provided that the halo selection function imposed by
DEEP2 galaxy selection can be understood in detail.  In addition, it
would be necessary to carefully account for the increased scatter in the
$M$--$\siggal$ relation imposed by selection effects.  
We describe a computational approach to achieving this in the
Appendix.

\section{Criteria for group-finder optimization}
\label{sec:methods}

\subsection{Group-finding terminology and success criteria}
\label{sec:criteria}
The aim of our group-finding exercise is to identify sets of galaxies
that are gravitationally bound to one another in common dark-matter
halos.  A perfect group catalog would identify
all sets of galaxies that share common halos and classify them all as
independent groups, with no contamination from other galaxies, and no
halo members missed.  Any realistic algorithm for finding
groups in a galaxy catalog, however, is subject to various sources of
error that cannot be fully avoided, owing largely to incompleteness in
the catalog and ultimately to the noise inherent in any discrete
process \citep{SzSz96}. 
Any individual type of error can typically be reduced to some extent
by varying the parameters of the group-finder, but this often comes at
the expense of increases in other kinds of error.  The classic example
of this is the trade-off between merging neighboring small groups
together into spuriously large groups on the one hand and fragmenting
large groups into smaller subclumps on the other \citep{NW87}.

Because there are inevitably such trade-offs between various different
group-finding errors, it is important to define clearly the criteria
by which group-finding success is to be judged and the requirements
for an acceptable group catalog.  As discussed by G05, the optimal
balance between different types of error will depend on the particular
scientific purpose to be pursued by study of the groups.  In the
present study, our primary goal is to produce a group catalog that
accurately reconstructs the abundance of groups as a function of
redshift and velocity dispersion, $N(\sigma, z)$.  As discussed in
\citet{NMCD02}, such a catalog can be used to place constraints on
cosmological parameters.  Therefore, our optimal group catalog
will be the one that most accurately reconstructs $N(\sigma,z)$. 
 It is also of interest to use the group catalog for studies
of galaxy evolution in groups (\emph{e.g.}, \citealt{Gerke07a}) or of
the evolution of group scaling relations (\emph{e.g.},
\cite{Jeltema09}); a catalog that can be used for those purposes is a
secondary goal.  These two goals will drive our choice of metrics for
group-finding success in what follows.

\subsubsection{What is a group?}

In tests using mock catalogs, the ``true'' group catalog is known, and
we are using our group-finding algorithm to produce a ``recovered''
group catalog; this leads to potential ambiguity in the meaning of the
word \emph{group}.  To distinguish clearly between the two cases, we
adopt terminology similar to that employed by \citet{Koester07a}. For
the purposes of discussing group-finding in the mocks, a \emph{group}
is defined to be a set of two or more galaxies (the \emph{group
  members}) that are linked together by a group-finding algorithm.
Galaxies that are not part of any group are called \emph{field
  galaxies}.  By this definition, a group is not necessarily a
gravitationally bound system; rather it is exactly analogous to a
group in the real data.  By constrast, a \emph{halo}, for the purposes
of discussing group finding, is defined to be a set of galaxies in the
observed mock (the \emph{halo members}) that are all actually bound
gravitationally to the same dark-matter halo in the background
simulation \footnote{The assignment of mock galaxies to halos of course
  depends on the simulation, halo-finding, and mock-making algorithms
  we employ; we discuss this futher in the paper describing the mocks
  (Gerke et al. in prep.).  For the purposes of this study, though,
  the galaxy-halo assignment can be taken as ``truth'', since the
  choice of algorithms has already been made.}. It is possible to have
a halo that contains only a single galaxy; such galaxies (and
their host halos) are called \emph{isolated} and are analogous to
field galaxies in the group catalog.  By comparing the set of groups
to the set of non-isolated halos in the mock catalog, then, it will be
possible to judge the accuracy of the group-finder.

It will also be useful to distinguish between the \emph{intrinsic}
properties of halos (\eg, the total richness, or number of halo
members above some luminosity threshold), the \emph{observable}
properties of halos (\eg, the observable richness, or total number of
halo members that are in the mock catalog after DEEP2 selection has
been applied), and the \emph{observed} properties of groups (\eg, the
observed richness, or total number of group members).  Unless
otherwise specified, we will always discuss the properties of groups
and halos as computed using their member galaxies: for example, the
velocity dispersion of a halo will always be the dispersion of the halo
members' velocities, $\siggal$, rather than the dispersion of the dark-matter
particles, $\sigdm$, unless we explicitly specify that we are talking
about a dark-matter dispersion.

\subsubsection{Success and failure statistics: basic definitions}
\label{sec:stats_defs}
There are two primary modes of group-finding failure, for which we
will adopt the same terminology used in  G05.  \emph{Fragmentation}
occurs when a group contains a proper subset of the members of a given
halo, while \emph{overmerging} refers to a case in which a group's
members include members of more than one halo.  A special case of
overmerging involves isolated galaxies that are spuriously included in
a group; such galaxies are called \emph{interlopers}.  It is also
possible for fragmentation and overmerging to occur simultaneously, as
when a group contains proper subsets of several different halos.

Fragmentation and overmerging are generally likely to lead to a wide
diversity of errors when a group catalog is considered on an
object-by-object basis, so it will be useful to define a set of
statistics that summarize the overall quality of the catalog.  Here we
will adopt the statistics used in G05 (with one addition,
$f_\mathrm{noniso}$), which can be summarized as follows.  On a
galaxy-by-galaxy level, we define the \emph{galaxy success rate}
$S_\mathrm{gal}$ to be the fraction of non-isolated halo members that are
identified as group members.  Conversely, the \emph{interloper
  fraction} $f_\mathrm{int}$ is the fraction of identified group members that
are actually isolated galaxies.  It is also worth considering the
quality of the field galaxy population, since a perfect group finder
would leave behind a clean sample of isolated galaxies.  We therefore
also compile the \emph{non-isolated fraction} $f_\mathrm{noniso}$, which is
the fraction of field galaxies that are actually non-isolated halo
members.  On the level of groups and halos, we define two different
statistics.  Broadly speaking, the \emph{completeness} $C$ of a group
catalog is the fraction of non-isolated halos that are detected as
groups, while the \emph{purity} $P$ is the fraction of groups that
correspond to non-isolated halos.  In general, the classic trade-offs
inherent in group-finding are evident in these statistics: changes to
the group finder that improve completeness or galaxy success will typically
have negative effects on purity and interloper fraction.

Attentive readers will notice here that we have not yet defined what
it means for a halo to be ``detected'' or for a group to
``correspond'' to a halo, so the meanings of of the terms
\emph{completeness} and \emph{purity} are still unclear.  These
definitions, which are somewhat subtle, are the
subject of the following sections.

\subsubsection{Matching groups and halos}
\label{sec:matching}

In order to compute the completeness and purity of a group catalog we
must first determine a means for drawing associations between groups
and halos.  In the case of groups identified in a mock galaxy catalog,
the most natural way to do this is consider the overlap between the
groups' and halos' members.  This basic approach has been used with
good success in many previous studies (\eg, \citealt{Eke04}, G05,
\citealt{Koester07a, Knobel09, Cucciati10, Soares-Santos11}).  We
associate each group to the non-isolated halo that contains a
plurality of its members, if any such halo exists (otherwise the
cluster is a false detection).  Similarly, we associate each
non-isolated halo to the group that contains a plurality of its
members (again if any such group exists).  In the case of ties,
\emph{e.g.}, when two halos contribute an equal number of galaxies to
a group (an example of overmerging), we choose the object that
contains the largest total number of galaxies, or, if this is still
not unique, the one with the largest observed velocity
dispersion\footnote{we would choose randomly if both tie-breaker
  criteria failed, although this does not occur in practice}.
Hereafter, we will use the term \emph{Largest Associated Object (LAO)}
to refer to the group (halo) that contains the plurality of a given
halo's (group's) members.

This matching procedure is rather lenient
and is by no means unique: a group can in principle be associated to a
halo with which it shares only a single galaxy, multiple groups can be
matched to the same halo (and vice-versa), and a cluster may be
associated to a halo that is itself associated to some other cluster
For example, if a halo $H$ with five members is divided into two
groups, $G_1$ with three members and $G_2$ with two, then $G_1$ and
$G_2$ are both associated to $H$, but $H$ is only associated to
the larger of the two groups, $G_1$ (see Figure 4 of G05 or Fig. 3 of
\citet{Knobel09} for depictions of other complicated associations).
This example also illustrates the difference between \emph{one-way}
and \emph{two-way} associations: $G_1$ is associated with $H$, and
vice-versa, so this is a two-way match; however, $G_2$ is
associated with $H$, but the reverse is not true, so this is a one-way
match.  

In G05, we used a more stringent matching criterion that made an
association only when the LAO contained more than 50\% of the galaxies
in a given group or halo.  This definition has the virtue of removing
the need to break ties between possible LAOs, but it is somewhat
problematic in the case of low-richness systems.  If, for example, a
halo containing four galaxies had two of its members assigned to the
same group by the group-finder, with the other two being called
field galaxies, the G05
criterion would class the group as a successful detection but would
deem the halo to be undetected.  Because of situations like this, we
choose here to separate questions of simple group \emph{detection}
from issues of group-finding \emph{accuracy}.  In order to assess the
latter, we also compute the overall \emph{matching fraction} $f$ of
each group-halo association: the fraction of galaxies in a given
system (group or halo) that are contained in its LAO.  In what
follows, we will use this fraction to consider more and less stringent limits
on accuracy when computing completeness and purity statistics.

\subsubsection{Purity and completeness}
\label{sec:purcomp}

To compute purity and completeness, it will be necessary to define the
criteria by which a group-halo association constitutes a ``good''
match, to be counted toward these statistics.  In general we will
count associations above some threshold in $f$, and we will compute
separate purity and completeness values for one-way and two way
matches.  We will represent these various purity and completeness
statistics using the symbols $^wP_f$ and $^wC_f$, where we are
only counting associations with match fractions larger than $f$, and $w=1$ or
$w=2$ indicates that we are counting one-way or two-way associations.

The simplest statistics to use are $^1P_0$ and $^1C_0$, which denote
the fraction of groups and halos, respectively, that have any
associated object whatsoever, regardless of match fraction or match
reciprocity.  These values are good for getting an overall sense of
the group-finder's success at making bare detections of halos, but
their usefulness is somewhat limited since, for example, one could
achieve $^1C_0 = ^1P_0 = 1$ simply by placing all galaxies into a
single enormous group (in this case, all halos would be associated to
the group, and the group would be associated to the largest halo).  A
more useful pair of statistics is $^2C_0$ and $^2P_0$, the fractions
of halos and groups that have \emph{two-way} associations, regardless
of match fraction.  These tell us the fraction of halos that were
detected without being merged with a larger halo and the fraction of
groups that are not lesser subsets of a fragmented halo.  In the
pathological all-inclusive cluster example above, $^2P_0 = 1$, but
$^2C_0$ is near zero, indicating a problem.  

This also illustrates the usefulness of comparing one-way
and two-way completeness and purity statistics for diagnosing problems
with a group finder.  If $^1C_0$ is substantially larger than $^2C_0$,
for example, then a significant fraction of detected halos must have
been merged into larger systems, so overmerging is a significant
problem.  Conversely, if $^1P_0$ is much larger than $^2P_0$, then
there must be substantial fragmentation in the recovered catalog.  It
will also be interesting to consider completeness and purity
statistics using different values for $f$, such as $^2C_{50}$ and
$^2P_{50}$, which were used in G05.  As discussed above, however, using
more stringent matching-fraction thresholds can give an overly
pessimistic impression of the overall detection success.  For our main
assessment of overall completeness and purity, then, we will use
$^2C_0$ and $^2P_0$, since these statistics use the broadest possible
definition of a ``good'' match that does not count fragments
and overmergers (beyond the largest object in each fragmented or
overmerged system) as successes.

\subsubsection{The velocity function of groups}

In addition to considering the detection efficiency of the group
finder on a system-by-system basis, for some science applications one
may also be interested in various properties of the group catalog as a
whole.  In the case of DEEP2, it has been shown \citep{NMCD02} that
the bivariate distribution of groups as a function of redshift and
velocity dispersion, $dN(z)/d\sigmav$, can be used to constrain
cosmological parameters, since it depends on the volume element $V(z)$
and on the evolving group velocity function $dn(z)/d\sigmav$, both of
which depend on cosmology.  \citet{MDNC02} and G05 have shown
previously that the VDM groupfinder can accurately reconstruct this
distribution in high-redshift spectroscopic surveys. 

In this study, we will use the reconstuction of the velocity function
as a second measure of group-finding success.  After we have optimized
the completeness and purity of our groupfinder, we will further
optimize the group-finder to reconstruct $dN(z)/d\sigmav$ as well as
is possible without sacrificing completeness or purity.  In practice,
this boils down to comparing the number counts of groups and halos in
bins of $z$ and $\sigmav$.  Since the distribution is quite steep in
$\sigmav$, it will be important to take some care in our choice of
binning.  We discuss these details below in
Section~\ref{sec:optimize_vf}

\section{The group-finding algorithm}
\label{sec:VDM}
\subsection{The Voronoi-Delaunay group finder}

The Voronoi-Delaunay method (VDM) group finder is an algorithm for
detecting groups of galaxies in redshift space from spectroscopic
survey data.  It has advantages over the usual Friends-of-Friends
(FoF) approach in very sparsely sampled datasets, when the linking
lengths required for FoF group-finding become larger than typical
group sizes (for a more detailed discussion of this point, see G05).
VDM makes use of the local density information that is obtained by
computing the three-dimensional Voronoi tesselation and Delaunay mesh
of the galaxies in redshift space.  The Voronoi tesselation is a
unique partitioning of space about a particular set of points (the
galaxies in this case), in which each point is assigned to the unique
polyhedral volume of space (the \emph{Voronoi cell}) that is closer to itself
than to any other point.  The Delaunay mesh is the geometrical dual of
the Voronoi tesselation and consists of a network of line segments
that link each point  to the points in immediately
adjacent Voronoi cells.  Galaxies that are directly linked by the
Delaunay mesh are called \emph{first-order Delaunay neighbors},
neighbors of neighbors are \emph{second-order Delaunay neighbors}, and
so on.

The VDM algorithm was first
described by \citet{MDNC02}, who showed that it could be used to
detect galaxy groups in a DEEP2-like survey.  In particular, they
showed that the VDM algorithm could be tuned to accurately reconstruct
the distribution of groups as a function of velocity dispersion
$\sigmav$ and redshift $z$, above some threshold in $\sigmav$; this
was confirmed by G05, who produced a preliminary DEEP2 group catalog
using a version the VDM algorithm.  VDM has also been applied
successfully to the VVDS \citep{Cucciati10} and zCOSMOS
\citep{Knobel09} redshift catalogs.  Readers are referred to G05 and
\citet{MDNC02} for detailed descriptions of the algorithm we will be
using in this study.  Here, we summarize the basic algorithm and the
differences from the version we used in G05.

After computing the Voronoi tesselation and Delaunay Mesh for a given
galaxy sample, the VDM algorithm proceeds in three phases.  In Phase
I, the galaxies are first sorted in increasing order of their Voronoi
cell volume, a time-saving step which ensures that group-finding is
attempted in very dense regions first.  Then, proceeding through this
sorted list in order, we consider each galaxy in turn as a ``seed''
galaxy for a galaxy group, provided that it has not already been
assigned to a group.  A cylinder\footnote{All VDM cylinder dimensions
  are \emph{comoving} distances and are converted to angular and
  redshift separations by assuming a flat $\Lambda$CDM cosmology with
  $\Omega_M=0.3$.  This cosmology is assumed regardless of the true
  background cosmology when running on mock catalogs, since it is what
  we assume when running on the DEEP2 dataset, to allow consistency
  with previous DEEP2 studies, particularly G05. It is straightforward
  to rescale the cylinder dimensions to different assumed background
  cosmologies.} is drawn around each seed galaxy with radius \rmin and
length $2\mathcal{L}_{min}$.  If that cylinder contains any
first-order Delaunay neighbors of the seed galaxy, they are deemed to
be part of a group, and the algorithm proceeds to Phase II.  If no
first-order neighbors are found in the Phase I cylinder, no group is
detected, and the algorithm proceeds to the next galaxy in the list.

In Phase II, a larger cylinder is defined around the seed galaxy, with
radius $\mathcal{R}_\mathrm{II}$ and length $2\mathcal{L}_\mathrm{II}$.  We count
the number of galaxies in this cylinder that are first or second-order
Delaunay neighbors of the seed galaxy, denoting this number by
$N_\mathrm{II}$.  Since the number density of observed galaxies varies with
redshift, we correct $N_\mathrm{II}$ by the ratio of number density of DEEP2
galaxies at $z=0.8$ to the local number density at the group redshift.
The number density is computed by smoothing the DEEP2 redshift
distribution and dividing by the comoving cosmological volume element.

The corrected value, $N_\mathrm{II}^{\mathrm{corr}}$, is taken as an
initial estimate of the size of the group and is used to scale the
final search cylinder in Phase III.  The Phase III cylinder is
centered on the barycenter of the Phase II galaxies and has radius
$\mathcal{R}_\mathrm{III} = \max(r\times(N_\mathrm{II}^\mathrm{corr})^{1/3},
\mathcal{R}_{min})$ and length $\mathcal{N}_\mathrm{III} =
\max(\ell\times(N_\mathrm{II}^\mathrm{corr})^{1/3}, \mathcal{L}_{min})$,
with $r$ and $\ell$ being the Phase III parameters of the
algorithm. All galaxies that fall within the Phase III cylinder are
deemed to be members of the group.  The algorithm then continues to
the next galaxy in the list that has not yet been assigned to a group
and repeats the procedure.

The VDM thus has six tuneable parameters (two for the search cylinder
in each of the three phases) that must be optimized for a particular
survey.  These are not fully independent, however.  For example, an
increase in the size of the Phase II cylinder will increase the
typical $N_\mathrm{II}$ values and so can be offset by a decrease in the
Phase III $r$ and $\ell$ parameters.  Furthermore, our group-finding
exercise (indeed, any group-finding exercise) can be conceptually
subdivided into two steps: \emph{group detection}, which occurs in
Phase I alone, and \emph{membership assignment}, which occurs in
Phases II and III.  The parameters that control each of those steps
can be tuned more or less independently of one another on the way to
determining an optimum set of group-finding parameters.

\subsection{Changes to the G05 Algorithm}
\label{sec:VDM_changes}
Before we leave discussion of the VDM algorithm, it is important to
make note of a few minor changes that we have made to the VDM
algorithm we used in G05.  First, we have used a redshift of
$0.8$ as a reference for correcting $N_\mathrm{II}$, since
$z=0.8$ is near the peak of the DEEP2 redshift distribution, in
contrast to the G05 reference value, $z=0.7$, where the redshift
distribution is rising sharply in the main DEEP2 sample.

We also made some important changes to the membership-assignment part
of the algorithm.  In G05 each group included all galaxies identified
in either Phase II or Phase III of the VDM algorithm, regardless of
whether or not the Phase III cylinder was larger than the Phase II
cylinder.  This meant that the Phase II cylinder dimensions had to be
kept relatively small, so as not to swamp small groups with interloper
field galaxies.  In testing the VDM on our new mock catalogs, we found
that this led to significant fragmentation of larger groups: the Phase
II cylinder was too small to accurately estimate their richnesses, so
the Phase III cylinder was significantly too small to include all
their members.

To some degree, this is unavoidable in a sparsely sampled survey, but
we found that we were able to mitigate it by allowing the Phase II cylinder to be
quite large, similar in scale to a massive cluster.  To gain this
advantage while avoiding problems in smaller groups, we decided not to
include Phase II galaxies in the final group memberships.  That is, we
use the Phase II cylinder to get a rough estimate of the number of
galaxies in the group by drawing a cylinder that is typically too
large and will pick up all the group members and possibly some field
galaxies.  The scaled Phase III cylinder then refines this estimate
and will frequently select only a subset of the Phase II galaxies for
the final group.  In practice, with a very large Phase II cylinder,
the $N_\mathrm{II}$ counts often simply include \emph{all} second-order
Delaunay neighbors, with the cylinder simply setting a maximum
distance at which such neighbors will be considered.  For this reason,
we find that varying the Phase II cylinder at relatively large sizes
has negligible impact on our results.  We thus focus mainly on optimizing
the Phase I and III parameters in what follows.

\subsection{Considerations for the EGS}
\label{sec:VDM_EGS}

Because the galaxies targeted in the EGS cover a very broad redshift
range with a fixed apparent magnitude limit, the range of galaxy
luminosities being probed varies dramatically from low to high
redshift, with only very bright ($L\ga L^\ast$) galaxies being observed
at $z\ga 1$ but extremely faint dwarfs included at low redshift.  The
presence of these introduces some complications into the group-finding
process.  The first has to do with the simple definition of a
``group.'' In the main DEEP2 sample, groups are systems containing on
the order of a few Milky-Way-sized galaxies at least.  At low $z$ in the EGS,
by contrast, we will also be capable of detecting systems consisting
of a single Milky-Way-sized galaxy and a few dwarfs similar to the
Magellanic Clouds.  Arguably we should not categorize the
latter systems as groups at all.  

More importantly, the faint, low-$z$ dwarfs present a challenge for
optimizing the VDM group-finder.  Phases I and II of the VDM algorithm
search for galaxies that are connected to a given seed galaxy by one
or two links in the Delaunay mesh.  Using Delaunay connectedness in
this way as a means of detecting groups of bright galaxies rests on
the assumption that group members of similar luminosity are likely to
be Delaunay neighbors.  When the much galaxies are included, this
assumption may break down, since dwarfs are much more numerous than
galaxies near $L^\ast$, and so it is possible that a bright galaxy's
local Delaunay mesh may be ``saturated'' by dwarfs, cutting off any
links to neighboring bright objects and preventing detection of the
larger group.  Indeed, in our initial experiments with mock EGS 
catalogs, we found that it was impossible to achieve satisfactory
performance with the VDM group-finder at both low and high redshift
simultaneously if the entire EGS galaxy sample was used.

If we choose to focus our group-finding efforts on systems containing
multiple bright galaxies, as in the main DEEP2 sample, then
fortunately there is a simple way of addressing both of the above
issues by limiting Phases I and II of the group finding to bright
galaxies only.  In particular, when computing the Voronoi partition
and Delaunay mesh in the EGS, we can restrict the low-redshift
($z<0.8$) sample to only those galaxies that are luminous enough that
they could have been observed at $z>0.8$.  To do this, we follow the
procedures used in \citet{Gerke07a}, who defined a set of diagonal cuts
in the DEEP2 rest-frame (\ie, $k$-corrected as in \citealt{Willmer06})
color-magnitude space, which correspond to the DEEP2 $R=24.1$ apparent
magnitude limit at different redshifts (see Figure 2 of that paper).
If we define such a cut that traces the faint-end limit of DEEP2
galaxies in color-magnitude space at $z=0.8$, we can then select only
galaxies brighter than this limit at lower $z$; these are the
low-redshift analogs of the main DEEP2 sample. (When performing this
selection, we also evolve the cut toward fainter magnitudes at lower
redshifts, according to the evolution of $L^\ast$ that was obtained in
\citealt{Faber07}, namely a linear evolution of 1.2 magnitudes per unit $z$).

For EGS groupfinding, we apply this
selection to the $z<0.8$ galaxy population before computing the
Voronoi and Delaunay information, and we consider only the selected
galaxies in Phases I and II of the algorithm.  This means that only
systems containing at least two bright galaxies (that would be
observable at $z\ge 0.8$) will be counted as groups.  In Phase III,
however, we consider all galaxies regardless of luminosity, since this
final membership-assignment step simply counts all galaxies in the
Phase III cylinder, without reference to the Delaunay mesh.  This
approach to group-finding in the EGS has the virtue of ensuring that
the groups in the EGS have similar selection, while also counting
dwarf members of the groups where they have been observed.

\section{Optimization on DEEP2 mock catalogs}
\label{sec:optimize}

Tthe VDM algorithm has six free parameters whose optimal values are
not immediately obvious.  It is thus very important to test the
algorithm on simulated data that reproduce the properties of the real
data as accurately as possible.  As discussed above in
Section~\ref{sec:mocks}, the mock catalogs developed in Gerke et
al. (in prep.) accurately reproduce a wide array of the observed
properties of the DEEP2 catalog, including color-dependent selection
effects that might disproportionately impact galaxies in groups
relative to those in the field.  Testing the VDM groupfinder on these
mocks will thus represent a significant improvement over the
group-finding effort in G05, which made use of mocks that lacked such
color-dependent effects.  The current mocks also have been constructed
for three different cosmological background models, one of which, for
the Bolshoi simulation, is very close to the model that best fits
current data.

In practice, we optimize the group-finding parameters by running the
VDM group finder repeatedly on the mock DEEP2 observational fields,
allowing the group-finding parameters to vary over a wide range in the
six-dimensional parameter space, and looking for parameter sets that
meet our optimization criteria.  Since the Bolshoi simulation
cosmology is in the best agreement with present data, we use this
simulation to perform the main optimization.  However, we repeated
this procedure on each of the three different sets of mock catalogs
described in Section~\ref{sec:mocks} (see
Table~\ref{tab:mockcosmo}) to test whether and to what degree the
optimal parameter set depends on the background cosmology.

For our purposes, the optimal set of group-finding parameters will be
the one that most accurately reconstructs the velocity function, as
measured using the velocities of the observed galaxies, while also
stiking the best possible balance between the purity and completeness
of the group catalog.  It is not immediately obvious that all of these
requirements can be met simultaneously within the six-dimensional VDM
parameter space.  However, the steps in the VDM algorithm divide
rather cleanly into a group-detection step (Phase I) and a
membership-assignment step (Phases II and III).  Since our purity and
completeness statistics are mostly a test of group-finding success,
whereas velocity dispersion measurements depend on assigning the right
galaxies to the right groups, it is reasonable to supose that the two
success criteria may be optimized at least semi-independently.
Indeed, experimentation reveals that completeness and purity are only
weakly coupled to the shape of the recovered velocity function, at
least near the optimum of the purity and completeness values: here,
purity and completeness depend mostly on the Phase I parameters of
VDM, while the velocity function reconstruction is mainly governed by
Phase III.

In the following, then, we optimize purity and completeness first and
then consider the velocity function.  Additionally, as in G05, we identify a
\emph{high-purity} parameter set, for which the purity of the catalog
is nearly maximized, at the expense of completeness.  We will use this
when constructing the DEEP2 and EGS group catalogs to identify a 
subset of groups that should be considered higher-confidence
detections than the rest.

\subsection{Purity and completeness}
\label{sec:optimize_pc}

\begin{figure*}
\epsfig{width=7in, file=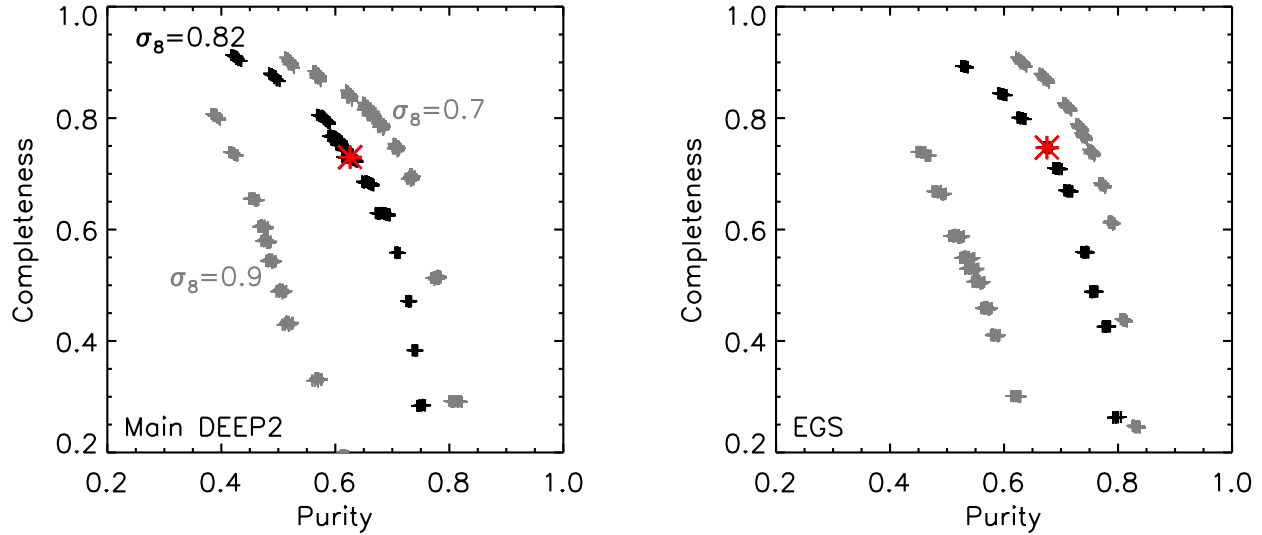}
\caption{Purity and completeness statistics for group catalogs
  computed with the VDM group-finder over a wide range of the
  algorithm's parameter space, for each of the three different mock
  catalogs discussed in Section~\ref{sec:mocks}.  The left panel shows
  results for mocks selected like the main DEEP2 sample, and the right
  panel shows results for the EGS mocks.  Each point in the plots
  shows the $^2P_0$ and $^2C_0$ statistics (as defined in
  Section~\ref{sec:purcomp}, and computed in aggregate over all mock
  lightcones) for a particular choice of group-finding parameters in a
  particular mock catalog.  The mock catalogs are identified by the
  value of $\sigma_8$ assumed for each one in the left panel (see
  Table~\ref{tab:mockcosmo}); the relative positions of values for
  each of the three mocks are the same in the right-hand diagram. The
  well-known trade-off between completeness and purity is evident for
  each mock catalog.  There are also substantial differences between
  the purity and completeness values measured in the different mock
  catalogs. }
\label{fig:purcomp}
\end{figure*}

Figure~\ref{fig:purcomp} shows the purity and completeness statistics
$^2P_0$ and $^2C_0$ that we obtained for widely varying choices
group-finding parameters in each of the different mock cosmologies.
Each data point in the Figure represents the completeness and purity
values (computed over all mock lightcones for each cosmology) that we
obtained for a given set of group-finding parameters and mock
cosmology.  Results are shown for both the main DEEP2 mocks and the
EGS mock catalogs.  A diagram like this is a very useful visualization tool
for group-finder optimization; it is similar in spirit to Figure 4 of
\citet{Knobel09}. The fundamental trade-off between completeness and
purity is readily apparent in the Figure: an increase in completeness
is always accompanied by a decrease in purity, and vice-versa.

For each mock cosmology depicted in the Figure, distinct clusters of
datapoints are apparent.  In all cases, each individual cluster
corresponds to a different value of the Phase I parameter \rmin; the
impact of varying the other five VDM parameters (rather widely in many
cases) is confined to the area of each cluster of points.  (The range
of \rmin values considered in the figure is $0.1\le $\rmin$\le 0.5$ Mpc.)
The obvious conclusion is that, for the purposes of optimizing simple
group \emph{detection}, \rmin is by far the most important parameter,
with all other parameters having a comparatively negligible effect on
detection efficiency.  In general, increasing \rmin improves the
completeness statistic while degrading the purity (and vice-versa).
It is apparent from the Figure, however, that improvements in either
completeness and purity are eventually subject to diminishing returns:
at sufficiently low values of \rmin, for example, completeness drops
rapidly, while purity remains approximately constant.  This fact sets
a practical range of interest for \rmin (roughly between 0.15 and 0.35
Mpc), beyond which changes in that parameter only serve to degrade the
quality of the catalog.  Within this range, the optimal choice of
\rmin is debatable, but a reasonable choice (also used by
\citealt{Knobel09}) is to take the one that gives a result in
purity-completeness space that is near the minimum distance from the
point $(1,1)$ that is obtained over the full parameter space.  We find
that this optimum is roughly in the range $0.225\la \mathcal{R}_{min}
\la 0.3$ Mpc (the red star denotes $\mathcal{R}_{min} = 0.25$ for the
main DEEP2 plot and $0.3$ for the EGS plot).  Since we will also find
below that the velocity function reconstruction depends on \rmin, we
will allow this parameter to vary slightly in the next step.

The current analysis already allows us to partially identify our
high-purity parameter set.  Since the purity of the catalog
effectively saturates at \rmin$=0.15$, and since purity is not
sensitive to any other parameters, we will be able to identify a
high-purity subset of the final group catalogs by setting \rmin to
0.15 and holding all other parameters fixed at their optimum values,
whatever those turn out to be.

Another notable feature of Figure~\ref{fig:purcomp} is the
significantly different purity and completeness values we obtained for
the different mocks.  This implies that our inferred success
statistics have a strong dependence on cosmology and particularly on
the level of clustering in each dark-matter simulation.  This can
plausibly be explained as follows.  The abundance-matching technique
used to construct the mocks requires, by construction, that the
luminosity function must match the one that is measured in the DEEP2
data by placing galaxies in halos and subhalos at fixed number
density.  As discussed in Gerke et al. (in prep.), as the clustering
amplitude $\sigma_8$ increases, so does the number density of halos at
fixed mass; hence, the abundance-matching algorithm places
\emph{fainter} galaxies in halos (and subhalos) of a given mass.  This
means in particular that massive halos will contain fewer galaxies
above a given luminosity---and thus fewer observable galaxies---as
$\sigma_8$ increases.  This will make these groups more difficult to
detect, since their observable galaxy populations will be sparser.
For example, the typical halo with two observed members will be more
massive in a more clustered cosmology, so its observed members will
typically be more widely separated in redshift space.  At the same
time, increasing $\sigma_8$ enhances the clustering of the isolated
background galaxies, making them more likely to be erroneously grouped
together as false detections.  The net
effect is that, when the luminosity function is held fixed, increasing
the galaxy clustering and halo mass function (\eg, by raising
$\sigma_8$ and $\Omega_M$) causes a decrease in \emph{both}
completeness and purity.

This result has important implications for the optimization of group
finders generally.  Since mock catalogs are usually constructed to
reproduce the observed galaxy luminosity function reasonably well,
this effect is likely to be generically present in mocks with
different background cosmologies and not just in catalogs produced
using abundance matching.  When assessing group-finders, then, it will
be very important to construct mocks using simulations whose
background cosmology are consistent with the current best-fit
cosmological parameters.  

Mock catalogs based on semi-analytic galaxy formation models applied
to the Millennium Simulation \citep{Millennium} have the dual
advantages of matching a wide variety of observed properties of the
galaxy population and of being easy to obtain and use, so they have
been widely used to test and optimize high-redshift group-finders.
For example, \citet{Knobel09} used the mock catalogs from
\citet{KW07}, and \citet{Cucciati10} constructed mocks using the
semi-analytic models of \citet{DB07} and the lightcone-construction
techniques of \citet{Blaizot05}.  However, the Millennium Simulation
had a background cosmology with $\sigma_8=0.9$, well above the
currently preferred value of $\sim 0.8$. Our tests here on mocks with
different cosmologies suggest that group-finders that were calibrated
on the Millennium mocks may have significant inaccuracies in their
estimated purity and completeness statistics.  (A similar statement
can be made about our earlier group-finding efforts in G05, for which
this paper should be considered a replacement.)  More generally, our
results suggest that group catalogs that were calibrated on mocks with
disfavored cosmologies should be treated with caution.

When summarizing the success and failure statistics of our
DEEP2 group catalog, then, we will use values computed using the
Bolshoi mock catalogs only.  For the remainder of this Section, we will
also focus our group-finder optimization efforts on those mocks.

\subsection{The group velocity function}  
\label{sec:optimize_vf}

\begin{figure}
\epsfig{width=3.25in, file=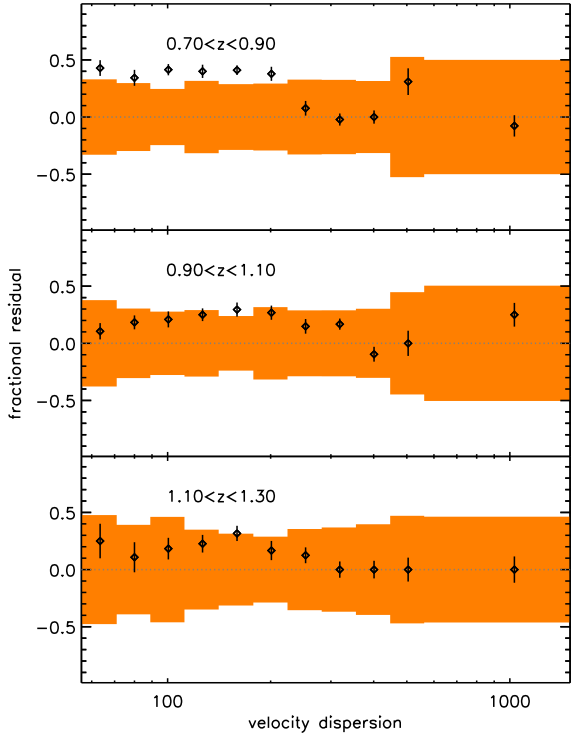}
\caption{Fractional error in the recovered group velocity function for
  our optimal set of DEEP2 group-finding parameters
  (Table~\ref{tab:parameters}).  Data points show the median fractional
  difference between the group and halo number counts (defined
  precisely in the text) in bins of $\sigmav$ and $z$, for the 40
  Bolshoi mock lightcones. Error bars show the standard error computed
over the 40 mocks.  The shaded regions indicate the fractional scatter
in the halo number counts (due to both sample variance and Poisson
noise).  For $\sigmav \ga 300$\kms, the fractional errors are
consistent with zero and/or are smaller than the sample variance for
all redshifts. At lower velocity dispersions, the group abundance is
systematically overestimated.}
\label{fig:velfn_resid}
\end{figure}

While maintaining a reasonable balance between completeness and purity
of the catalog, we also wish, by tuning the VDM parameters, to produce
a catalog of groups that accurately reconstructs the distribution
function of observed velocity dispersions for halos in the mock,
$dN/d\sigmav$, at all redshifts of interest.   More specifically,  we
will focus on the high-$\sigmav$ end of the velocity function, since
this region of the distribution is exponentially sensitive to
changes in cosmology, and we ultimately wish to use our group catalog
for cosmological tests.  In G05, we found that VDM could reconstruct
the velocity function of the \citet{YWC04} mocks accurately at
$\sigmav\ge 350$\kms; here we will also endeavor to reconstruct the
velocity function above a threshold value of $\sigmav$ that is as low
as possible.

Throughout this study, when we discuss the velocity function of groups
or halos, we are talking about the measured dispersion $\siggal$ of
the member galaxies' line-of-sight peculiar velocities (though we will
generally drop the superscript for brevity).  Since the overwhelming
majority of DEEP2 groups will contain only a few galaxies, we make use
of the so-called gapper algorithm to measure $\sigmav$.  The gapper
measures velocity dispersion using the gaps between the measured
line-of-sight velocities $v_i$ in a given group, after the $v_i$ have
been sorted in ascending order:
\begin{equation}
\sigma_G = \frac{\sqrt{\pi}}{N(N-1)}\sum_{i=1}^{N-1}i(N-i)(v_{i+1}-v_i). 
\label{eqn:gapper}
\end{equation}
This has been shown to be the most 
robust of several possible estimators in the limit of small samples
\citep{BFG90}.

In practice, testing the velocity-function reconstruction amounts to
comparing the number counts of halos and reconstructed groups in the
mock catalogs, in bins of $\sigmav$ and $z$.  In a perfect
group-finder, these histograms would be exactly equal.  To assess the
successfulness of the VDM algorithm, we will use a similar approach to
the one we used in G05: we compare the fractional error in the
recovered histogram to the fractional size of the field-to-field
dispersion in the mock histograms.  The field-to-field dispersion
serves as a proxy for the sample variance (sometimes called cosmic
variance) in the group number counts.  As a rough rule of thumb, where
the error in reconstructing the velocity function is smaller than the
sample variance (and also not systematically high or low over a wide
range in $\sigmav$), we take the reconstruction to be acceptable,
since the measurement error is subdominant to the irreducible
uncertainty in the measured velocity function that arises from
sampling a finite volume of space.  We will endeavor to reduce this
error as much as possible, however, focusing mainly on the
high-dispersion end of the velocity function, since it is the most
sensitive probe of cosmology.

It will be important to be careful in our choice of binning when
performing this test.  Because the velocity function of groups is
quite steep at the high-$\sigmav$ end, and because the volume of
DEEP2 is relatively small, there will be very few groups at large
values of $\sigmav$, especially at high redshift.  If we choose a
binning in $\sigmav$ that is too narrow, then most high $\sigmav$
bins will contain zero or one group, and a slight inaccuracy in the
measured velocity dispersion of a given group could lead to very
large apparent fractional errors, when in fact the group detection
introduces only a minor inaccuracy in the velocity function---one that
would have a negligible effect on the inferred cosmology.  We would
therefore like to use a coarse enough binning to ensure that every bin
is expected to contain at least a handful of groups.  Since changes
in cosmology affect the \emph{shape} of the velocity function as well
as the normalization, though, we would also like to choose a fine
enough binning to capture at least some of the shape information.  

Somewhat fortuitously, this set of requirements is identical to the
one we would use to select a binning for performing cosmological tests
with the velocity function: we wish to measure the shape of the
velocity function as well as possible given our dataset, but standard
techniques for constructing likelihood functions over the cosmological
parameters assume that the dataset contains at least a few objects per
bin \citep{HC06}.  Hence, for the purposes of testing the VDM
velocity-function reconstruction, we will choose a binning that would
be appropriate for using the resulting group catalog in cosmological
tests, with at least a few groups falling in each bin.  Experimenting
with the halo population in the mock catalogs, we find that a
reasonable choice is to construct even bins in $\log_{10}\sigmav$, of
width 0.1, with the addition of a single, broad bin covering all
values above $\log_{10}\sigmav = 2.75$, with redshift bins of constant
width 0.2.  Because G05 found that VDM
can accurately reconstruct the velocity function above a threshold of
$\sigmav=350$\kms, and lacking any other compelling reason to specify
a particular positioning for our bin edges, we choose to arrange our
logarithmic bins such that one of the edges falls near the G05
threshold, at $\log_{10}\sigmav = 2.55$.  

Within these bins, we compute the fractional difference between the
counts of halos and groups and compare it to the fractional sample
variance in these counts, as described above.  By performing this
procedure for group catalogs computed over a wide range in the VDM
parameter space, we can search for an optimum set of parameters that
minimizes the error in the reconstructed velocity function at high
$\sigmav$.  Figure~\ref{fig:velfn_resid} shows the reconstruction
error obtained for this optimum set in the Bolshoi mock catalogs.  The
black points denote the median fractional difference between the
binned number counts of groups and halos,
$(N_{groups}-N_{halos})/N_{halos}$, and the error bars show the
standard error in this quantity, with the median\footnote{We consider
  the \emph{median} fractional error in this diagram, rather than the
  mean, because the former statistic is more resistant to the outlier
  values we occasionally encounter.  Sample variance and shot noise
  mean that a few of the Bolshoi mock lightcones contain few (in one
  case zero) halos in one or more of our bins, despite the fact that
  the typical (\ie, median) lightcone contains at least five halos in
  each bin.  However, the group-finder still tends to find at a few
  groups in all bins, which leads to a very large fractional error for
  a few lightcones and strongly biases the mean.}  and standard error
computed over all forty Bolshoi mock lightcones.  The shaded regions
show the fractional sample variance in each bin, as well as indicating
the extent of each bin in $\sigmav$.

To arrive at the parameter set used in the figure, we varied the VDM
parameters over a wide range in parameter space (though constraining
\rmin to the narrower range of 0.225--0.3 identified in the previous
section).  We used the measured fractional reconstruction errors and
their uncertainties (\ie, the data points and error bars in the
Figure) to construct a statistic similar to $\chi^2$ for the bins with
$\sigmav > 350$ \kms.  We then gave detailed consideration to
parameter sets near the minimum in this statistic, tuning the
parameters by hand to find the parameter set that gives an error in
the high-$\sigmav$ velocity function that is smaller than the sample
variance in all bins and is not systematically biased to high or low
values.  These optimum parameters are listed in
Table~\ref{tab:parameters}.  This set of parameters is also indicated
by the red asterisk in Figure~\ref{fig:purcomp}.  It is notable in
Figure~\ref{fig:velfn_resid} that this
parameter set gives a fractional reconstruction error near
zero above a threshold of $\sigmav\approx 300$ \kms, a slight
improvement over the value of $350$ \kms we achieved in G05.  At lower
dispersions, the velocity function is overestimated by the
group-finder at all redshifts and this bias is larger than the sample
variance at low $z$.

\begin{deluxetable}{lccc}
\tabletypesize{\small}
\tablewidth{0pt}
\tablecaption{VDM group-finding parameters used for the different
  sample in this study}
\tablehead{
\colhead{Parameter}\tablenotemark{a}&\colhead{Main}&\colhead{EGS}&\colhead{High-Purity}
}
\startdata
\rmin  & 0.25 & 0.3 & 0.15  \\
$\mathcal{L}_{min}$  & 10.0   & $\cdots$ \tablenotemark{b}   & $\cdots$ \\
$\mathcal{R}_\mathrm{II}$   & 0.8   & $\cdots$  & $\cdots$ \\
$\mathcal{L}_\mathrm{II}$   & 8.0   & $\cdots$  & $\cdots$ \\
$r$                 & 0.225 & $\cdots$  & $\cdots$\\
$\ell$              & 10.5  & $\cdots$  & $\cdots$
\tablenotetext{a}{All values are given in comoving $h^{-1}$ Mpc,
  assuming a flat $\Lambda$CDM cosmology with $\Omega_M = 0.3$.}
\tablenotetext{b}{Values not listed are the same as used for the main
  DEEP2 sample.}
\label{tab:parameters}
\end{deluxetable}

While performing the optimization, we found that the reconstructed
velocity function was sensitive to the Phase I and Phase III VDM
parameters only; the Phase II parameters had no noticeable impact over
the range we considered.  As discussed in
Section~\ref{sec:VDM_changes}, this is because we have allowed the
Phase II cylinder to take very large values, such that we are simply
counting all second-order Delaunay neighbors in Phase II in nearly all
cases.  This suggests that the Phase II cylinder serves little
practical purpose in our implementation of the VDM group-finder, where
we have allowed it to be very large.  However, keeping it in place has
no negative impact on the recovered catalog, and it has the virtue of
ensuring that we never count very distant neighbors in Phase II.  On
the other hand, if we allowed the Phase II cylinder to take on smaller
values the resulting decreases in $\mathcal{N}_\mathrm{II}$ could be offset
by increases in the Phase III parameters, so that our parameter space
for optimization would still be effectively reduced to four
parameters, rather than six.

\subsection{The Extended Groth Strip} 
\label{sec:testing-egs}

Because DEEP2 spectroscopic targets in the EGS were selected
differently from those in the main DEEP2 sample, it is reasonable to
expect that the optimal parameter set for group finding will also be
different.  We therefore repeat our parameter optimizaton procedure
separately on the mock lightcones constructed with EGS selection.
There are two primary differences between the EGS sample and the main
DEEP2 sample.  First is the presence of a low-$z$ sample in the EGS,
which includes very faint dwarf galaxies.  As discussed in
Section~\ref{sec:VDM_EGS}, we handle this by excluding from Phase I and II
galaxies fainter than those included in the main DEEP2 sample.    The
second difference is the somewhat denser sampling in the
EGS permitted by the use of a four-pass slitmask-tiling algorithm
instead of the two-pass approach used in the rest of DEEP2.

This higher sampling rate would naively be expected to make groups
easier to detect, and so one might initially suppose that
completeness, for example, will rise relative to the rest of DEEP2, at
fixed values of the Phase I \rmin parameter.  Recall, however, that
the Phase I step of VDM looks for first-order Delaunay neighbors, and
increasing the density of objects in the galaxy catalog will also
shorten the typical linking length in the Delaunay mesh.  Since the
Phase I cylinder has a long, narrow shape along the line of sight, it
is possible that the denser sampling will actually tend to reduce the
number of first-order neighbors within the cylinder in favor of nearer
neighbors that do not fall along the line of sight.  In this case, a
slightly larger value for \rmin might be indicated.   

In fact, we find
the latter situation to be the case: generally speaking, we need a
\emph{larger} value of \rmin to strike a
similar balance between between purity and completeness to the one we
achieved in the main DEEP2 sample.  In
Figure~\ref{fig:purcomp}, the red asterisks in both panels sit at approximately the
midway point between the maximum attainable purity and completeness
values.  However, in the EGS (right-hand panel) this point corresponds
to \rmin=0.3, somewhat larger than the value of 0.25 we used for the
main DEEP2 sample.   We will therefore use \rmin=0.3 when finding groups in
the EGS.  

One might also initially imagine that the Phase III parameters should be
reduced (for fixed Phase II parameters) to account for an expected
increase in $\mathcal{N}_\mathrm{II}$ values owing to the higher sampling
density.  However, since $\mathcal{N}_\mathrm{II}$ counts only second-order
Delaunay neighbors, it is far from clear that its value will be
strongly coupled to the sampling density.  Indeed, we find that, by
increasing \rmin to 0.3 and keeping \emph{all other parameters fixed},
we can obtain an acceptable reconstruction of the velocity function at
high redshifts, $z>0.7$.  At lower redshifts, the reconstructed
velocity function overestimates the true one slightly; however, the
volume probed here is quite small, and we do
not plan to use these groups as a cosmological probes.  In the
interest of constructing a group catalog that is as similar as
possible to the main DEEP2 sample, then, we opt to hold all VDM parameters
constant in the EGS with the exception of \rmin.

\section{The DEEP2 VDM group catalog}
\label{sec:catalog}
To construct a catalog of DEEP2 galaxy groups, we take all galaxies
that yielded good redshifts (DEEP2 quality flag 3 or 4).  We then
exclude galaxies at $z>1.5$, which are mainly high-redshift QSOs and
MgII absorbers.  For the main (non-EGS) DEEP2 sample, we also exclude
galaxies with $z<0.6$.  In the EGS, we exclude galaxies that are too
faint for observation at $z>0.8$ in Phases I and II of the VDM
procedure, as discussed in section~\ref{sec:testing-egs}.  This yields
a sample of 22,144 galaxies in the three high-redshift DEEP2 fields
and 12346 galaxies in the EGS.  Of these, 3100 high redshift galaxies
($14\%$ of the sample) and 3070 EGS galaxies ($25\%$ of the sample)
are assigned to groups by the VDM group-finder.  The group fraction in
EGS is naturally higher because of the higher sampling density in that
field.

\subsection{The catalogs of galaxy groups}

We ran the VDM algorithm on each of these samples using the parameters
listed in Table~\ref{tab:parameters}.  This produces two catalogs for
the main sample---one primary catalog and one high-purity
catalog---and two more for the EGS.   To identify the high-purity
subset of the main group catalog, we match the primary and high-purity
catalogs using the matching techniques described in
section~\ref{sec:matching}.  Groups in the main catalog that are the
LAOs of the high-purity groups are then flagged as members of the
high-purity catalog.  

The main catalog of VDM-identified DEEP2 galaxy groups consists of
1295 groups with $z>0.6$ in the main DEEP2 sample (outside the EGS)
and 1165 groups in the redshift range $0<z<1.5$ in the EGS.  Of these,
74\% of the main DEEP2 group catalog and 68\% of the EGS group catalog
is made up of systems with two observed members (\ie, galaxy pairs).
The catalog is presented in Table~\ref{tab:groupcatalog}, in which the
groups are organized by observational field and sorted by velocity
dispersion within each field.  Groups that were also detected in the
high-purity catalog are identified in the rightmost column of the
table.  These groups are substantially more likely than the rest to
correspond to real dark-matter halos (though they are not guaranteed
to do so, just as the other groups in the catalog are not necessarily
bogus).  Galaxies belonging to the groups are listed in
Table~\ref{tab:galcatalog}, with each galaxy's unique DEEP2
identification number listed alongside the ID number of its parent
group.  These catalogs are also available in electronic format from
the DEEP2 data release webpage\footnote{{\tt
    http://deep.berkeley.edu/dr4}}; they supersede the earlier DEEP2
group catalog presented in G05.

\begin{figure*}
\centering
\epsfig{width=6in, file=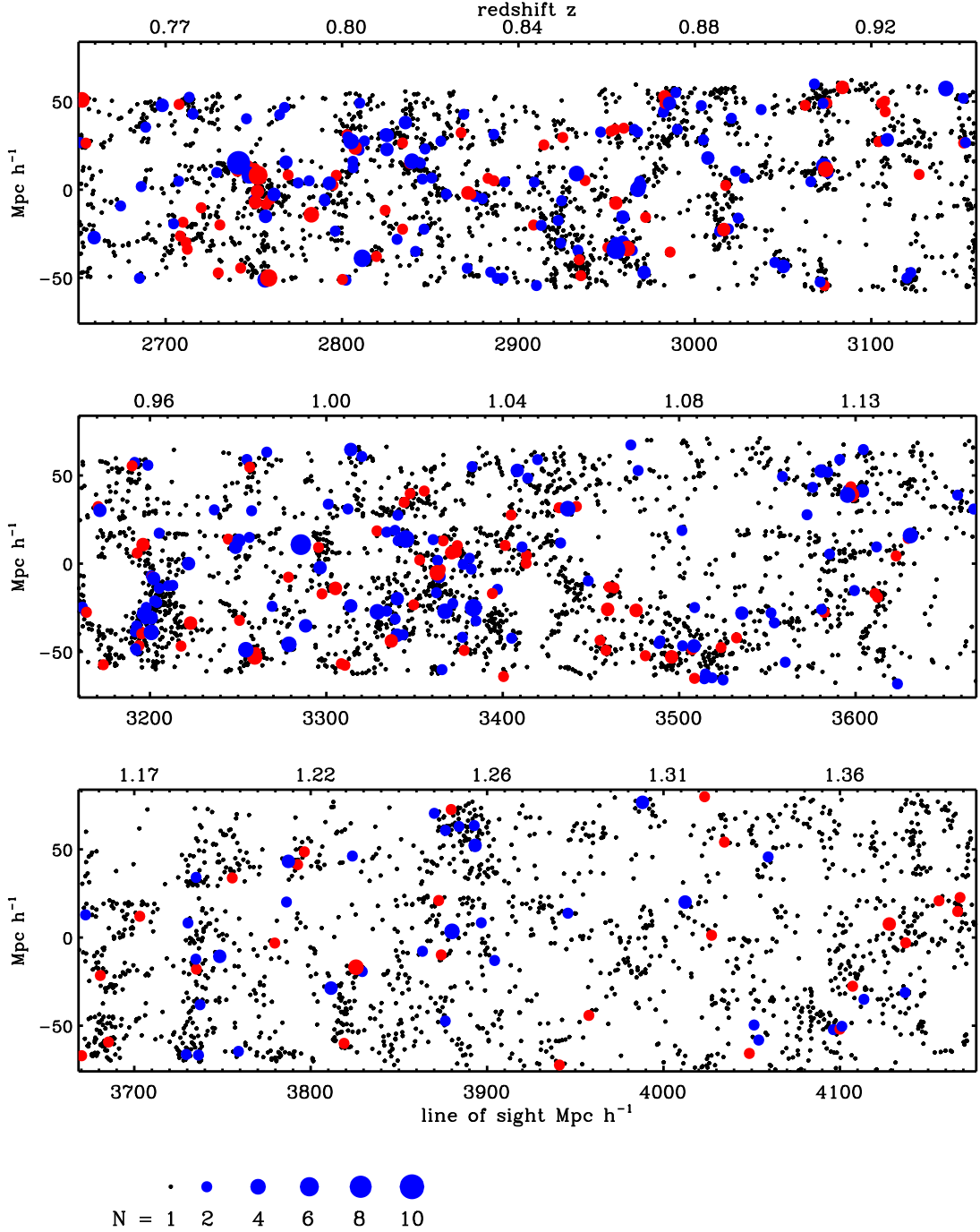}
\caption{The redshift-space distribution of
  groups and field galaxies in the largest of the three DEEP2 fields
  (DEEP2 field 3, at 23 hours RA).  Each panel shows a different
  redshift range, and the three panels taken together represent a
  contiguous volume covering the redshift range $0.75\le z\le 1.4$.
  The vertical axis in each panel shows transverse distance along the
  Right Ascension direction on the sky; the plots are projected along
  the Declination direction.  Black points show the positions of field
  galaxies, while red and blue circles show the positions of groups.
  Blue circles indicate groups that are part of the high-purity
  subsample, and red circles show other groups.  The size of each
  circle indicates the richness of the group. The horizontal empty
  region about two thirds of the way from the bottom of each panel is
  caused by a gap in the DEEP2 photometry.}
\label{fig:cone}
\end{figure*}

Figure~\ref{fig:cone} shows the redshift-space distribution of groups
in DEEP2 field 3 (at 23 hours R.A.), which is the DEEP2 field whose
spectroscopic observations cover the largest area on the sky
(approximately one square degree).  Black points in the figure show
the positions of field galaxies, while larger circles show groups,
with the size of the circle indicating the group richness.  Blue
circles are members of the high-purity sample; the rest of the
groups are indicated in red.  

\begin{figure}
\epsfig{width=3.5in, file=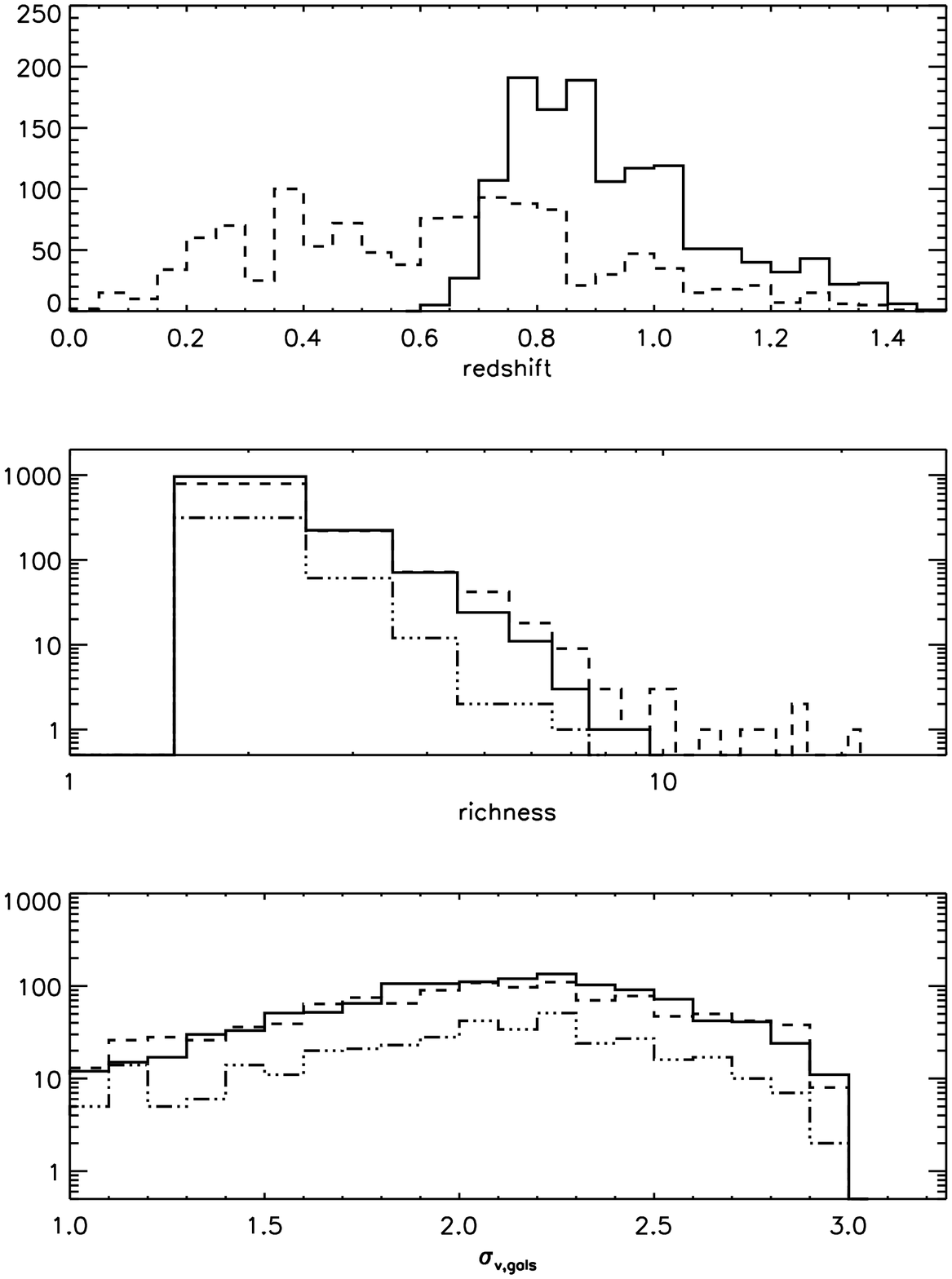}
\caption{Histograms of various properties of the DEEP2 group catalog.
  Solid lines show properties of groups in the main sample (excluding
  the EGS), and dashed lines show groups in the EGS.  Dot-dashed lines
show EGS groups at $z>0.75$ (for better comparison to the main DEEP2
sample).}
\label{fig:properties}
\end{figure}

Figure~\ref{fig:properties} shows distributions of various basic group
properties in the EGS and the main DEEP2 catalogs (redshift, richness,
and velocity dispersion).  To allow for a fairer comparison between
the EGS and the main DEEP2 sample, we have also plotted histograms for
the subsample of the EGS group catalog at $z\ge 0.7$.  It is
interesting to note that the high-redshift EGS sample contains roughly
half as many groups as the rest of DEEP2.  This is in spite of the
fact that the EGS covers only 20\% of the area covered by the main
DEEP2 sample.  The higher observed group abundance in EGS derives
partially from the higher sampling rate of galaxies (EGS contains 37\%
as many $z>0.7$ galaxies as the rest of DEEP2 combined, despite having
only a fifth of the area).  This means that a larger fraction of
low-mass halos will have two observed galaxies in the survey and thus
qualify as groups.   The higher density
of groups also arises partially from the larger value of \rmin we used
to detect groups in the EGS.  Our purity and completeness values have
remained roughly the same as in the rest of DEEP2; since the number of
observable halos has increased, this means that the number of false
detections must also have increased, which is a result of our larger
Phase I cylinder.

\subsection{Success and failure statistics of the catalogs}
\begin{deluxetable}{lcccc}
\tabletypesize{\small} \tablewidth{0pt} \tablecaption{Summary of the
  group-finding success statistics for the DEEP2 and EGS group
  catalogs} \tablehead{
  \colhead{Statistic\tablenotemark{a}}&\colhead{Main}& \colhead{Main
    hi-purity}&\colhead{EGS} & \colhead{EGS hi-purity} } \startdata
$^1C_0$ & 0.75 & 0.49 & 0.79 & 0.45 \\ $^2C_0$ & 0.72 & 0.46 & 0.74 &
0.41 \\ $^1C_{50}$ & 0.70 & 0.46& 0.68 & 0.39 \\ $^2C_{50}$ & 0.68 &
0.44& 0.66 & 0.38 \\ $^1P_0$ & 0.62 & 0.71 & 0.67 & 0.77 \\ $^2P_0$ &
0.61 & 0.71 & 0.65 & 0.76 \\ $^1P_{50}$ & 0.54 & 0.62 & 0.55 &
0.63\\ $^2P_{50}$ & 0.54 & 0.62 & 0.55 & 0.63\\ $S_\mathrm{gal}$ &
0.70 & $\cdots$\tablenotemark{b}& 0.66 & $\cdots$\\ $f_\mathrm{int}$ &
0.46 & $\cdots$&0.43 &$\cdots$\\ $f_\mathrm{noniso}$ & 0.04 &$\cdots$
&0.06 &$\cdots$\\ \tablenotetext{a}{Uncertainties in the purity and
  completeness values, as computed from the binomial distribution, are
  smaller than 1\% and are suppressed here for brevity.}
\tablenotetext{b}{Galaxy-level statistics are not reported for the
  high-purity catalogs, since these catalogs simply identify a subset
  of the groups in the main group catalogs.}
\label{tab:catalog_properties}
\end{deluxetable}

We have chosen to optimize our group-finding parameters to accurately
reconstruct the velocity function of groups while striking a balance
between the purity and completeness statistics $^2P_0$ and $^2C_0$ for
the catalog as a whole.  For the catalog we obtain, it will also be
useful to compile some of the other success and failure statistics
defined in Section~\ref{sec:criteria}, such as $S_\mathrm{gal}$ and
$f_\mathrm{int}$, both for the catalog as a whole
and as a function of redshift and group properties.  Many of the
success and failure statistics we will discuss in this section are
summarized in Table~\ref{tab:catalog_properties}.

\begin{figure*}
\centering
\epsfig{width=2.3in, file=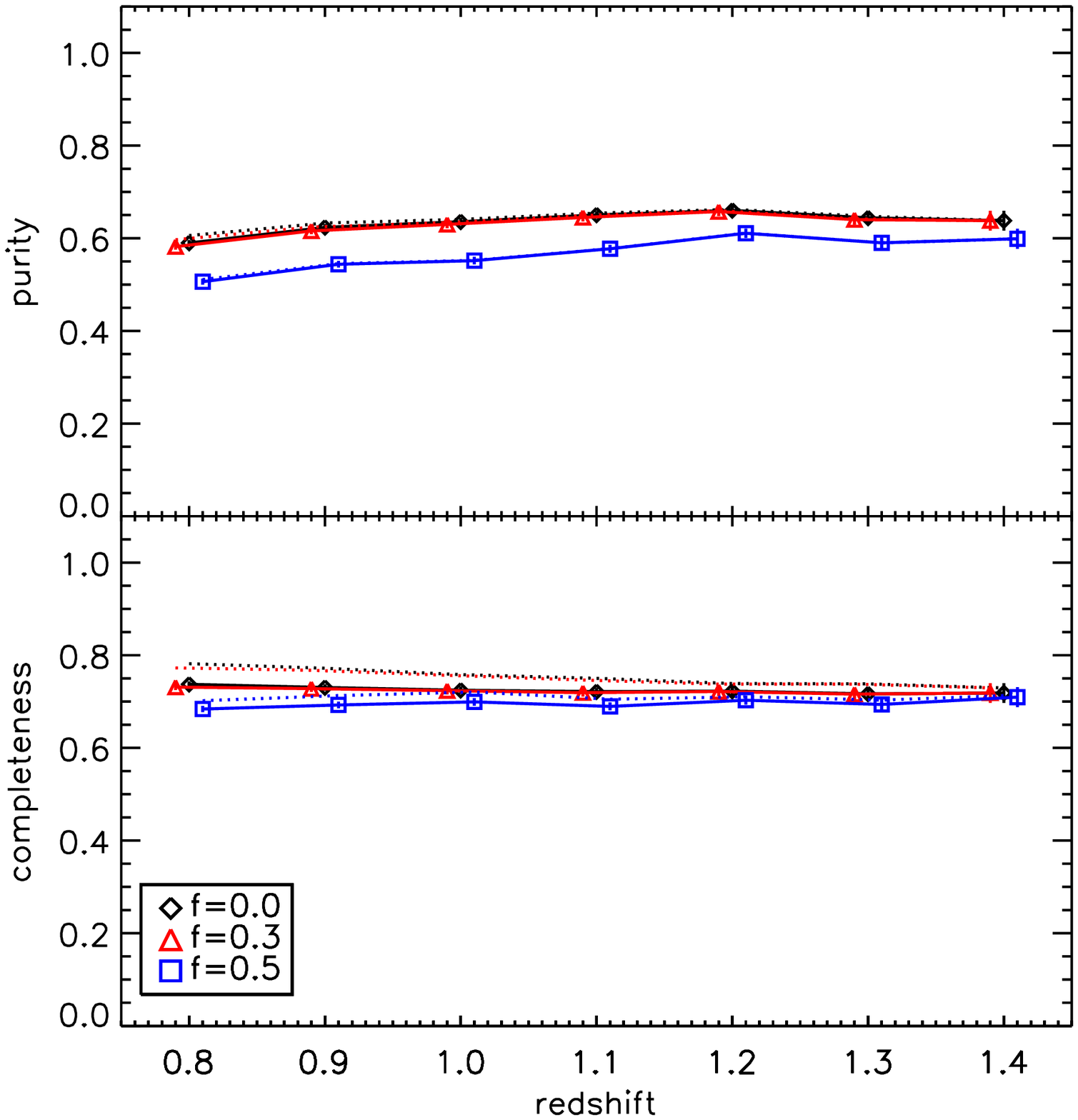}
\epsfig{width=2.3in, file=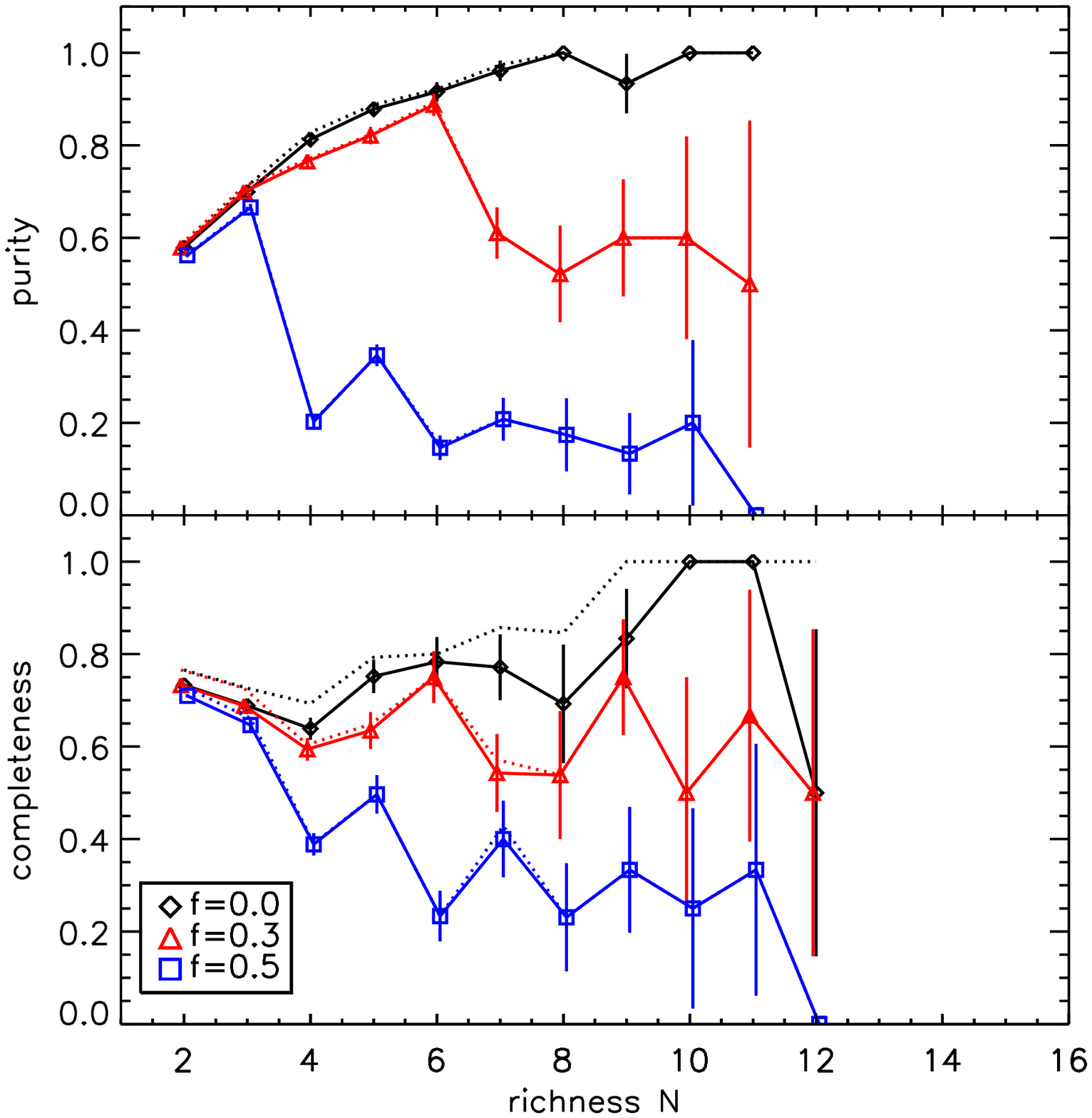}
\epsfig{width=2.3in, file=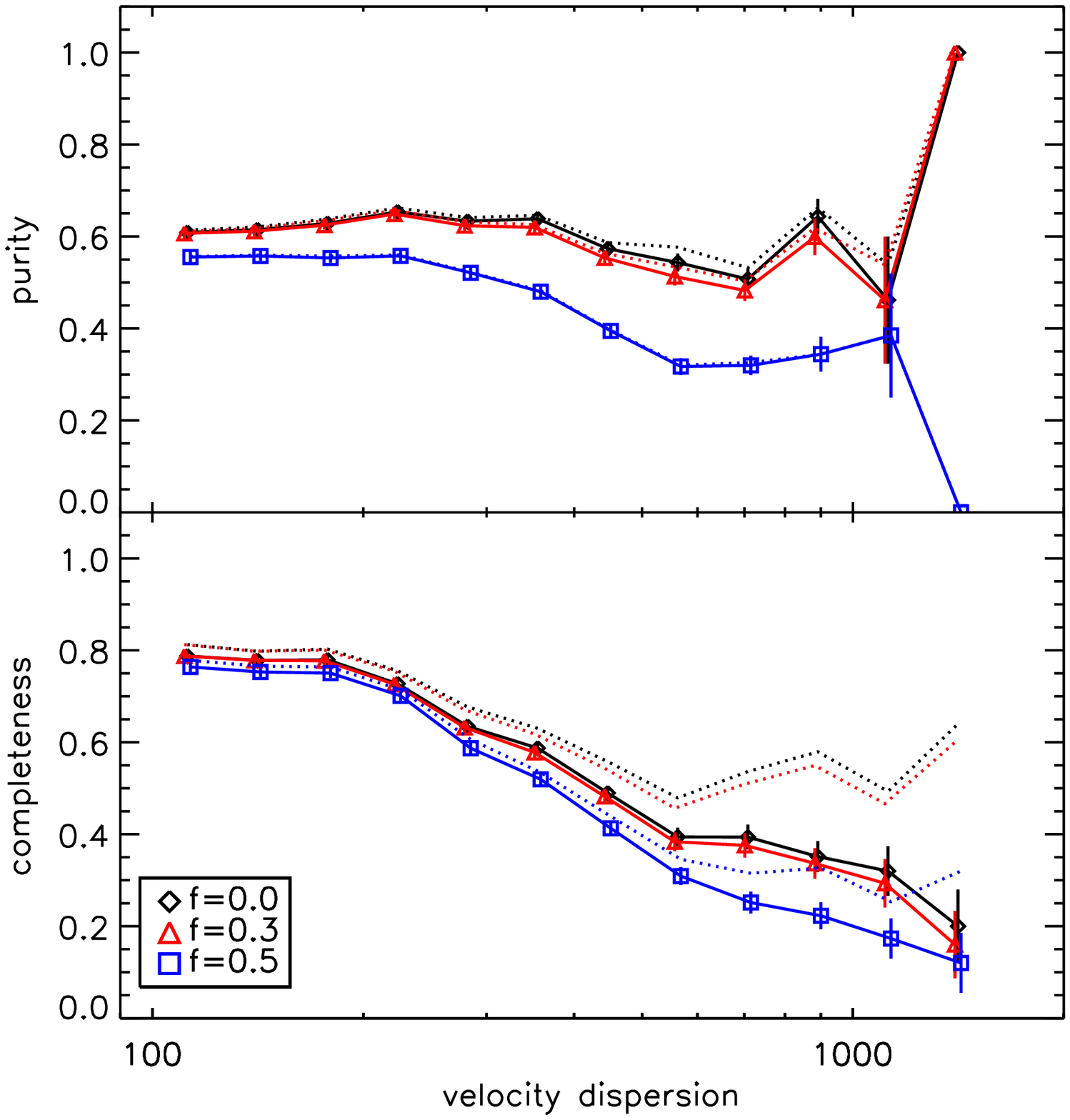}
\caption{Various purity and completeness statistics for the main DEEP2
  group sample, computed as a function of redshift, richness, and
  velocity dispersion. 
  Datapoints show the statistics $^wP_f$ and $^wC_f$ (defined
  Section~\ref{sec:purcomp}) for match reciprocity $w=1, 2$ and match
  fraction $f=0, 0.3, 0.5$,
  computed over all forty mock lightcones from the Bolshoi simulation.
  Error bars on these values are computed using the binomial
  distribution. One-way ($w=1$) statistics are shown as dotted lines and
  two-way statistics as solid lines.}
\label{fig:purcomp_nz}
\end{figure*}

\begin{figure}
\epsfig{width=3in, file=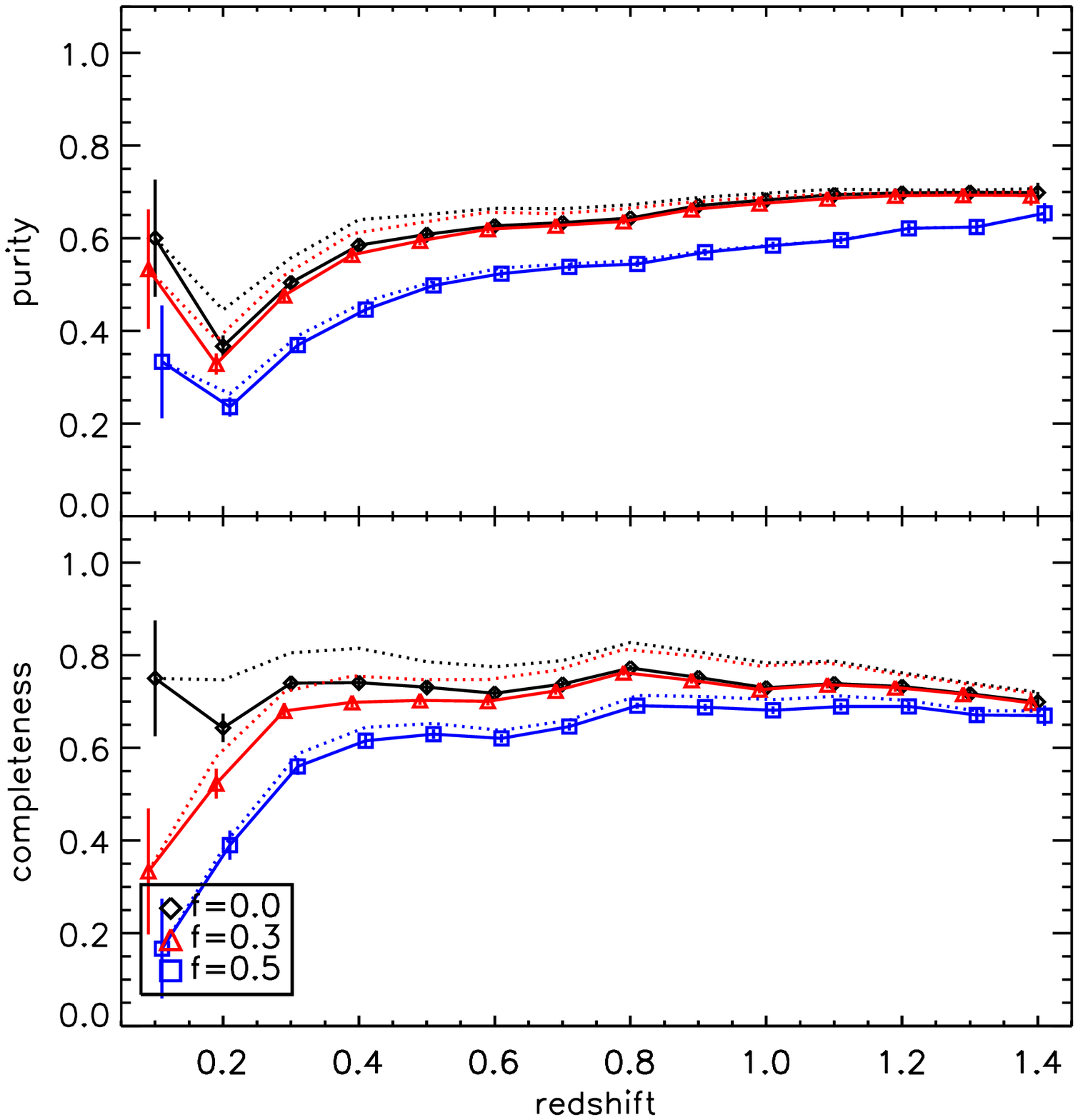}
\caption{Similar to the left-hand panel of
  Figure~\ref{fig:purcomp_nz}, but for the EGS mock catalogs.}
\label{fig:purcomp_z_egs}
\end{figure}

Figure~\ref{fig:purcomp_nz} shows six different definitions of both
purity and completeness: $^wP_f$ and $^wC_f$ for one-way and two-way
matching ($w=1$ and $2$) and matching fractions $f=0, 0.3$ and $0.5$.
Each statistic is shown as computed in bins of redshift, richness and
velocity dispersion.  It is notable that, for the catalog as a whole, the
different measures of purity and completeness are similar and are all
quite flat with redshift.  In particular, since the one-way and
two-way statistics are quite similar, we can conclude that the
fragmentation and overmerging rates are low for the catalog as a
whole.  However, when we look at the purity and completeness
statistics as a function of richness and velocity dispersion, it is clear that
group-finding errors become more prevalent for richer and higher
$\sigmav$  groups. 

For example, if we consider a stringent 50\% matching fraction, both
the completeness and purity of groups with $N > 5$ are quite low, even
though our usual $^2P_0$ and $^2C_0$ statistics are quite high.  We
also note that the $P_{50}$ and $C_{50}$ curves show an odd-even
``sawtooth'' pattern in the richness direction.  This is a
manifestation of of the weaknesses of these statistics that we
discussed in Section~\ref{sec:criteria}: a halo with $N=5$ needs three
detected members to be classified as a successful detection by the
$^2P_{50}$ standard, but a $N=6$ halo needs four detected members.
These plots suggest that, although a large fraction of high-richness
halos are successfully detected, many are fragmented and, conversely,
although almost none of the high-richness groups are false detections,
most of them are either overmergers, or have a large interloper
fraction, or both.  

This situation is unfortunate, but it is also
difficult to improve upon.  If we make a diagram similar to
Figure~\ref{fig:purcomp} showing $^2P_{50}$ versus $^2C_{50}$ for
groups with $N\ge 4$ only, for example, we can see a similar trade-off
between these two statistics, such that increases in one statistic are
generally offset by decreases in the other.  We have therefore chosen
not to further tune our parameters to improve the high-richness
success parameters.  Instead, we caution that although
high-richness DEEP2 groups can be treated as very high-confidence
\emph{detections}, their \emph{memberships} may be problematic.  It
may therefore be advisable, when considering individual high-richness
groups, to reconsider their memberships in a non-automated way, \eg,
as was done when considering X-ray detected EGS groups in
\citet{Jeltema09}.
 
The success as a function of velocity dispersion is potentially more
troubling, especially as regards completeness.  The purity is
relatively flat as a function of $\sigmav$, with a small drop around
$400$ \kms, which likely occurs because groups with a larger Phase III
cylinder are likely to be have a higher contamination by interlopers.
A stronger and more sustained drop in the completeness,\footnote{It is
  important to note that this result is quite different from the one
  shown in Figure~8 of G05, which showed that $^2C_{50}$ was flat with
  $\sigmav$.  As we were testing VDM for this study, we found that
  that figure suffered from a software error which erroneously
  calculated a flat completeness value.  In fact a similar dropoff in
  completeness was present in the G05 catalog.  Since the current
  study constitutes a replacement for G05, this error should not
  impact any future studies.} especially the two-way completeness,
starts at around $200$ \kms.  At higher values of $\sigmav$, the
one-way and two-way completeness values are quite different,
indicating that the pieces of the high-dispersion halos that VDM
detects are being merged with other (presumably lower-dispersion)
systems.  The drop in completeness can be explained by noting that the
halo population is dominated by systems with $N=2$ (\ie, pairs) at all
dispersions, and for these systems, $\sigmav$ is simply proportional
to the velocity difference between the members,
$\sigmav=\sqrt{\pi}\Delta v/2$ (see Equation~\ref{eqn:gapper}).
Detection in Phase I would require these two galaxies to be
first-order Delaunay neighbors, but as they become more widely
separated in redshift space, they become increasingly less likely to
be linked.  Furthermore, the Phase I cylinder half-length has an
effective length of $\sim 850$ \kms in velocity at $z\sim 1$, so pairs
with $\sigmav \ga 750$ \kms are completely undetectable in the absence
of interlopers\footnote{An example of such a system is a halo with
  $\sigdm = 500$\kms and two observed members with typical velocities
  that are oriented in opposite directions along the line of sight.}.

On the face of it, the rapid drop in completeness with $\sigmav$
raises concerns about using the VDM catalog to measure the abundance
of high-dispersion groups as a cosmological test.  The situation may
not be as bad as it initially appears, however.  As we just explained,
the halos that are likeliest to go undetected are systems with two
observed galaxies.  These will typically be halos at low mass whose
measured $\sigmav$ values have been artificially boosted by
small-number statistical effects.  Indeed, as shown in the lower right
panel of Figure~\ref{fig:dilution}, the population of halos above any
threshold in $\sigmav$ includes a very broad range of masses,
including many with $M\la 10^{13}M_\odot$, owing to the very large
scatter in velocity dispersion measured using the observed galaxies in
the halos.  Such low-mass systems that have been scattered to high
$\sigmav$ are contaminants in the high-mass sample that one would like
to construct for use in cosmological tests.

In the Appendix, we develop a theoretical approach to predicting the
velocity function of an observed group catalog as a function of
cosmological parameters, expanding the \citet{NMCD02} analysis to
include observational selection effects and the impacts of
contamination from low-mass systems with high measured dispersion
values.  Although it is essential to include these effects in the
theoretical predictions, it is not necessary that the actual
contamination in the group catalog arise from the exactly the same
low-mass halos that have large true velocity dispersions, as long as
the contamination in the group catalog is at the same level as
expected in the halo population.  Since our group catalog accurately
reproduces the halo velocity function at high $\sigmav$ (as shown in
Figure~\ref{fig:velfn_resid}), it appears that the contamination
levels at high dispersion in the halo and group popluations are
similar.  That is, the impurities in the catalog (which are also
mostly pairs) account for the undetected pairs at high dispersion.  It
may thus still be reasonable to use this group catalog for a test like
the one proposed in \citet{NMCD02}, although it will be important to
investigate the incompleteness and contamination of the group catalog
very carefully.  Such a detailed analysis of the suitability of our
group catalog for cosmological tests is beyond the scope of this study
though, and apart from a brief discussion in the Appendix, we defer it
to future work.

In the case of the EGS, we find similar (though noisier) trends of
purity and completeness with $N$ and $\sigmav$.  As shown in
Figure~\ref{fig:purcomp_z_egs}, however, the purity and completeness
statistics are roughly constant at high redshift, but they degrade
somewhat at the redshifts below those probed in the rest of DEEP2.
The likely reason for this has to do with the effective sampling rate
of bright group members in the low-redshift EGS.  As we discussed in
Section~\ref{sec:VDM_EGS}, when we perform VDM Phases I and II in this
regime, we consider only galaxies bright enough that they could have
been observed at $z\ge 0.8$.  The EGS target-selection downweights
low-redshift galaxies so that the redshift distribution of targets is
roughly constant, and since the low-redshift pool of targets also
includes faint galaxies that might be observed in lieu of bright ones,
the sampling rate of bright galaxies is sharply reduced at low
redshift relative to what can be attained at high redshift.  A more
sparsely sampled galaxy population is certain to result in a
lower-quality group catalog, since there are fewer observable halos
overall, and their members are more widely separated.  Thus, although
we have attempted to select groups as uniformly as possible, there are
limits to how well this can be accomplished.  For this reason, we
advise particular caution when comparing the low-redshift group
catalog in the EGS to the rest of DEEP2.

\begin{figure}
\centering
\epsfig{width=3in, file=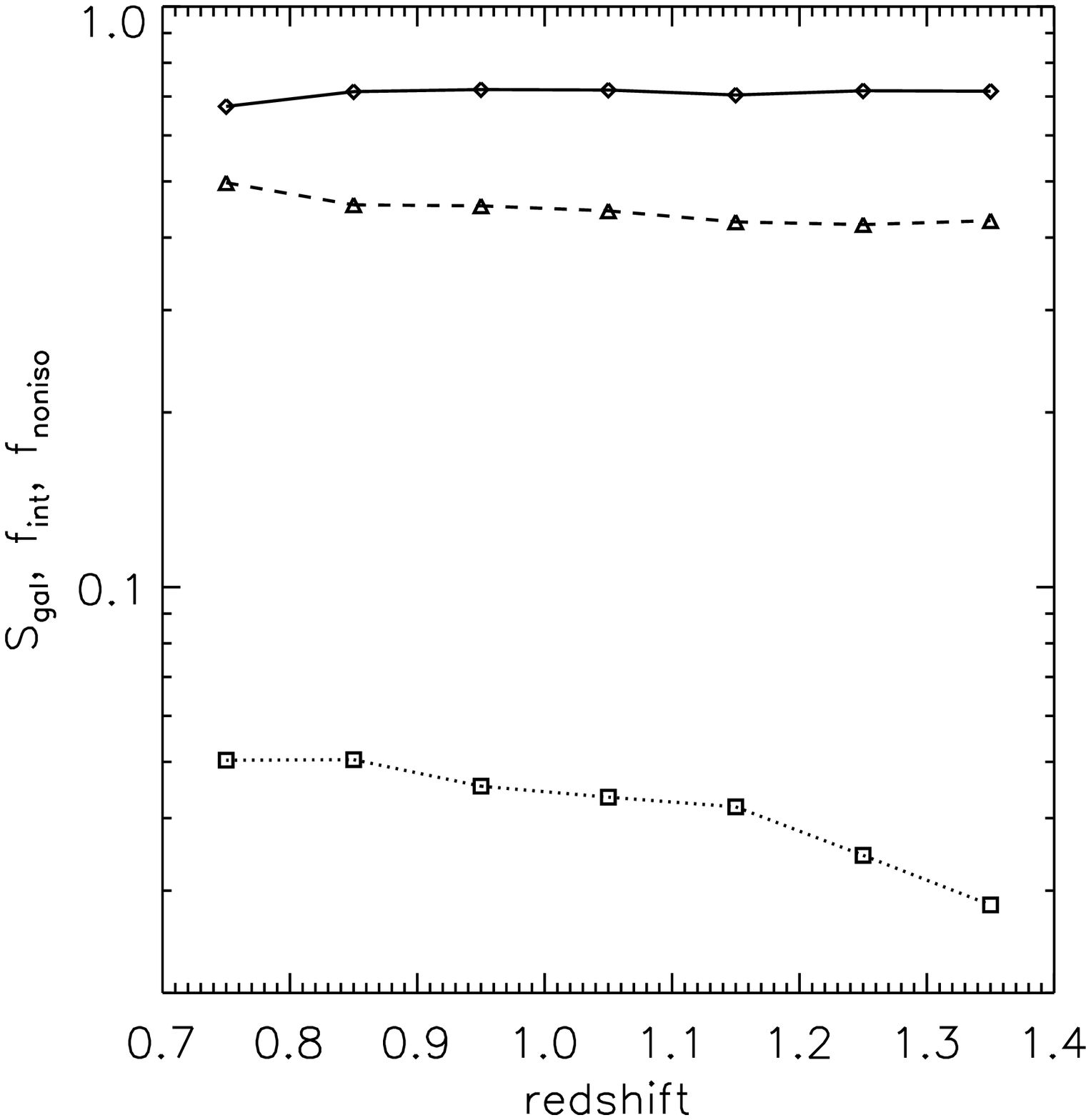}
\epsfig{width=3in, file=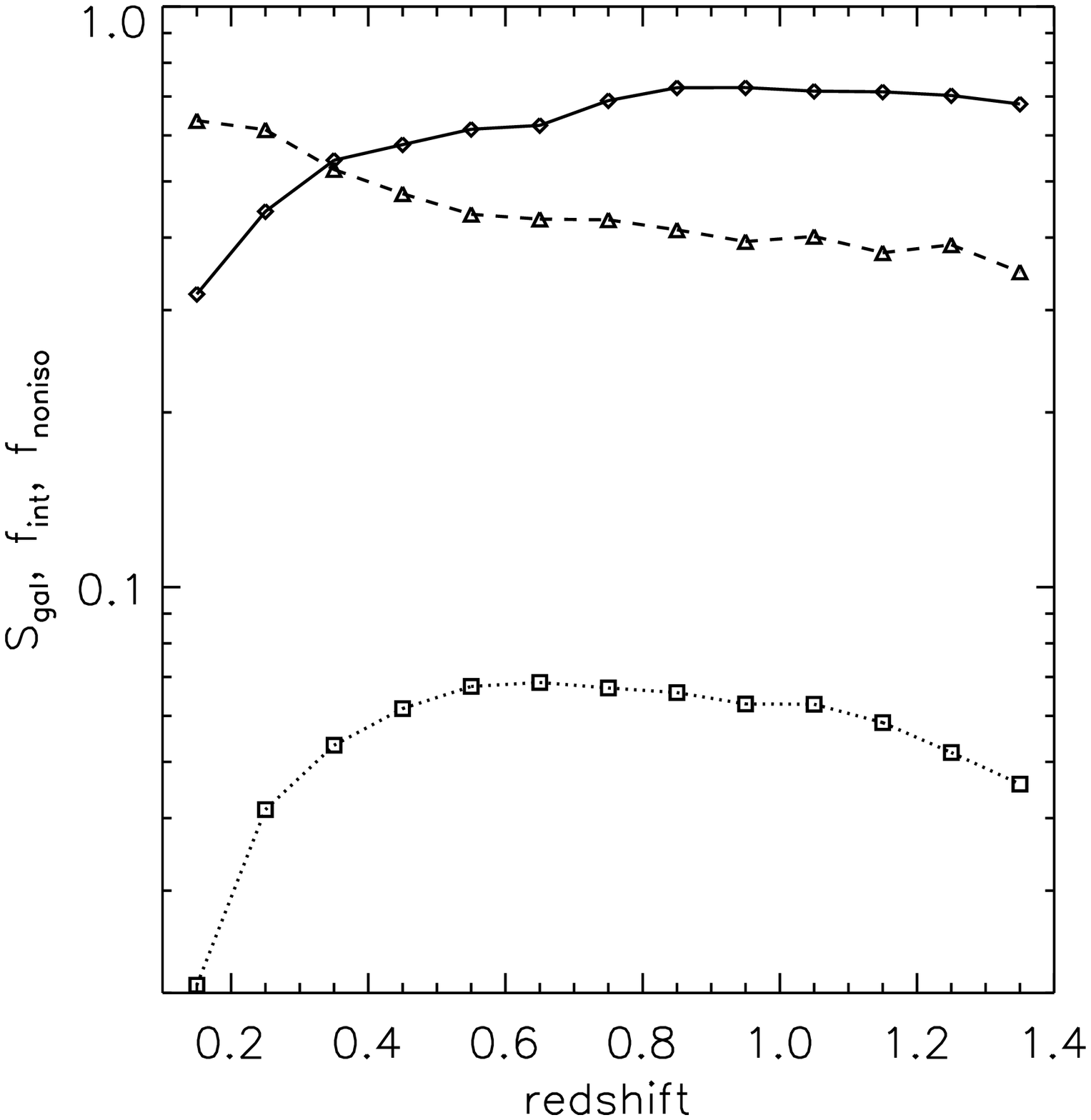}
\caption{Galaxy-level success and failure statistics (defined in
  Section~\ref{sec:stats_defs} for the main
  DEEP2 sample (top) and EGS (bottom), as a function of redshift.  The
  solid lines show the galaxy success rate, $S_\mathrm{gal}$; dashed lines
  show the interloper fraction, $f_\mathrm{int}$, and dotted lines show the
  nonisolated fraction, $f_\mathrm{noniso}$.}
\label{fig:galsuccess_z}
\end{figure}

In addition to identifying individual groups, the group catalog also
partitions the galaxy catalog into a set of group members and a set
of field galaxies.  Since galaxy groups and clusters are an
interesting environment for studying galaxy formation and evolution,
it will be worthwhile to consider the success and failure
statistics for the galaxy population that we defined in
Section~\ref{sec:criteria}.  These statistics, computed for the
catalogs as a whole, are summarized in Table~\ref{tab:catalog_properties}.
As shown in Figure~\ref{fig:galsuccess_z}, these values are
nearly constant with redshift at high $z$; however, there is a
significant worsening of the galaxy success statistics at low redshift
in the EGS.  The increase in interloper fraction at low redshift is
likely caused by the presence of faint isolated galaxies that happen
by chance to fall in the Phase III cylinder of some group (the Phase
III cylinder can be many tens of Mpc long in comoving line-of-sight
distance to account for the finger-of-god effect.)  The reduction in
$S_\mathrm{gal}$ likely derives from the lower sampling rate in the low-$z$
EGS that we discussed above.

Finally, we note that within the halo model for galaxy
formation, it is natural to divide the galaxy population into central
galaxies, which are located at the centers of distinct dark-matter
halos, and satellite galaxies, which are located in subhalos.
Considering the properties and evolution of centrals and satellites
separately has recently been a popular and fruitful approach to galaxy
evolution studes.
Ideally, one would hope that the population of field galaxies
identified by a group-finder would constitute a relatively pure
sample of central galaxies that could be studied separately from the
satellite population.  In tests on the DEEP2 mock catalogs, we find
that the fraction of field galaxies that are actually satellites is
$11\%$, which is a mild improvement over the $16\%$ satellite fraction
in the overall population of galaxies in the mock.  (These numbers are
$12\%$ and $18\%$, respectively, in the EGS mocks.)  The satellite
fraction can be higher than $f_\mathrm{noniso}$ since some truly isolated
galaxies in DEEP2 will be satellites if their central galaxies were
not scheduled for DEIMOS spectroscopy or did not yield good redshifts
(this also means that even a \emph{perfect} groupfinder would not
yield a satellite fraction of zero for the field galaxies).

\subsection{Comparison to other high-redshift spectroscopic group
  catalogs} 

There are two other spectroscopic galaxy surveys in the literature
that bear a particular resemblance to DEEP2 in targeting galaxies at
redshift of order unity with sample sizes on the order of tens of
thousands of galaxies.  These are the VIMOS VLT Deep Survey
\citep[VVDS;][]{VVDS} and zCOSMOS \citep{zCOSMOS} samples.  Group
catalogs have been computed for both of these samples using the VDM
group-finder by \citet{Cucciati10} (herefter C10) and \citet{Knobel09}
(hereafter K09),
respectively.  (The latter authors also computed a group catalog using
a friends-of-friends algorithm and used information from both
algorithms to construct a final group catalog for zCOSMOS.)  Since the
samples and methods used in those studies are broadly similar to ours
here, it will be useful to compare the DEEP2 group catalog with the
VVDS and zCOSMOS results.

The DEEP2 group catalog presented here is larger than the VVDS
and zCOSMOS catalogs, and it probes higher typical redshifts.  The
DEEP2 catalog contains 3016 groups in total,  compared
with 318 in VVDS and  1473 in zCOSMOS.  The
latter two catalogs lie primarily in the range $0.2<z<1.0$, with only
a few systems at higher $z$, whereas the DEEP2 catalog focuses on the
range $0.7<z<1.4$.  

There are important differences among the reported success and failure
statistics of the various catalogs.  As we do, C10 find that they can
accurately reconstruct the velocity function of groups for $\sigmav
\ga 350$\kms.  K09 do not consider the velocity function, but their
VDM catalog significantly overestimates the group abundance at all
richnesses, while their combined friends-of-friends and VDM catalog
accurately reconstructs the abundance.  C10 and K09 both report purity
and completeness values using the one-way and two-way $P_{50}$ and
$C_{50}$ statistics, so we can compare our catalog with theirs on
those terms.  We will focus on two-way statistics, since these are the
most stringent test of group-catalog quality and since the one-way and
two-way statistics for our catalog are quite similar.
Table~\ref{tab:catalog_properties} summarizes various purity and
completeness statistics for the DEEP2 catalog.  In terms of overall
purity and completeness, this catalog, with $(^2P_{50}, ^2C_{50}) =
(0.54, 0.68)$ is intermediate between the zCOSMOS $(0.72, 0.78)$
values and the VVDS values of $(0.49, 0.53)$.  This is not unexpected,
since the DEEP2 sampling rate is intermediate between the other two
surveys, and sampling density is the most important factor governing
group-finding success in redshift space.

When we consider the completeness and purity for \emph{high-richness}
groups, however, there are some rather striking differences between
the DEEP2 catalog and the other two.  As shown in
Fig~\ref{fig:purcomp_nz}, the DEEP2 $P_{50}$ and $C_{50}$ statistics
are rather low for groups and halos with more than three members.  By
contrast, both C10 and K09 report purity and completeness statistics
that are nearly flat with richness.  As we discussed above, we have
not been able to find a parameter set that increases the high-richness
purity or completeness without causing a decline in the other
paremter.  We have run VDM on the DEEP2 (Bolshoi) mock catalogs using
the optimal parameter sets determined in C10 and K09 and have found
that in both cases the high-richness completeness is substantially
improved, but this is at the expense of the purity.  In any event, we
note that the $^2P_0$ and $^2C_0$ statistics are quite high for our
high-richness halos and groups, so it is not correct to say that they
are mostly failures.  Instead, the membership assignment appears to be
somewhat problematic for these systems.

There are a few possible explanations for the differences between our
high-richness statistics and the others.  In the case of K09, it may
be possible to partially or fully account for the difference by the
fact that zCOSMOS has a higher sampling density than DEEP2.  However,
the difference is quite puzzling in the case of C10, since the VVDS
sampling rate is substantially lower than DEEP2's.  Another difference
that distinguishes both the C10 and the K09 catalogs from ours is
their redshift range covered and the resulting luminosity range
probed.  The main DEEP2 sample at $z>0.75$ only probes galaxies with
$L\ga L^\ast$ \citep{Willmer06}, so a DEEP2 group with five members,
say, has five \emph{bright} members and thus is quite massive.  With
similar flux limits to DEEP2, but lower redshift ranges, VVDS and
zCOSMOS will probe faint dwarf galaxies at low redshift, and only
brighter galaxies at high redshift, so that systems with five members
will be much less massive at $z=0.1$ than at $z=0.9$.  The presence of
the fainter galaxies in lower-mass systems means that the typical
effective sampling density of groups will be higher in VVDS and
zCOSMOS than it is in DEEP2 (recall that we exclude faint galaxies
from the group-detection step at low redshift in the EGS).  This may
make the detection of high-richness groups easier in the lower $z$
samples.

A final potential explanation for the differences in reported purity and
completeness statistics is the differences between the mock catalogs
used to estimate these values.  As we showed in
Section~\ref{sec:optimize_pc}, small differences in the background
cosmology of a mock catalog can lead to significant differences in the
measured purity and completeness statistics for the VDM groupfinder.
C10 and K09 both utilized mock catalogs
based on the Millennium simulation, whose cosmological parameters
(particularly the value $\sigma_8=0.9$) lie well outside the range
preferred by current data.  In the case of our mock catalogs, a higher
value of $\sigma_8$ leads to depressed purity and completeness
statistics, but since the Millennium mocks used a different
prescription for adding galaxies than we do, it is difficult to intuit
the effect that the background cosmology will have on the inferred
success and failure statistics.  Nevertheless, since properties of the
mock galaxies (\eg, the luminosity function) match observations, while
the halo population is not consistent with the favored cosmology, the
halo occupation statistics of galaxies are likely to be inaccurate.
One must therefore interpret purity and completeness statistics
computed from these mocks with some caution.  

In this study, we have made a substantial effort to use mocks (the
Bolshoi mocks) whose galaxy population and background cosmology are as
consistent as possible with a wide range of observations.  We have
also taken pains to ensure that we are striking the best possible
balance between completeness and purity given the stated goals of our
group-finding efforts.  Although the C10 and K09 catalogs nominally
appear to have better purity and completeness than ours (at high
richness and under very stringent definitions of purity and
completeness), given the differences between our mock catalogs and
theirs, it is nevertheless difficult to say whether this reflects a
true difference among the datasets or an insufficiently realistic set
of mocks in the earlier studies.

\section{Conclusion}
\label{sec:conclusions}

We have constructed and released to the public a catalog of galaxy
groups detected in redshift space from the spectroscopic galaxy sample
in the DEEP2 Galaxy Redshift Survey, including the Extended Groth
Strip.  Groups are detected using the Voronoi-Delaunay Method group
finder \citep{MDNC02}, as optimized on new, highly realistic mock
galaxy catalogs, using similar techniques to the ones we developed in
G05 for early DEEP2 data, using a previous generation of DEEP2 mocks.
The catalogs developed here replace those presented in G05.  Because
galaxies in the EGS region are selected with a different algorithm
than used in the rest of DEEP2, we perform a separate optimization of
the group finder on mock catalogs specially constructed to resemble
the EGS.  We also chose to exclude faint, low-redshift galaxies from
the first steps of group-finding in the EGS, to ensure that the
selection of the EGS group catalog is approximately uniform in
redshift and similar to the rest of DEEP2.

In the course of performing the optimizations, we developed a
generalized definition of the purity and completeness of the group
catalog, which avoids some problems that arise under the definitions
we used in G05.  We argue in particular that the G05 definition is
overly stringent and, by construction, gives results that are strongly
(and spuriously) richness dependent for low-richness systems.  Using
our more general purity and completeness measures, we find that our
optimized group catalog is 72\% complete and 61\% pure for the main
DEEP2 sample, and 74\% complete and 67\% pure for the EGS.  We also
construct a group catalog using parameters chosen to maximize purity
(at the expense of completeness), and we make note of these systems in
our main catalog.  The high-purity subsample is 71\% pure (76\% in
the EGS).  We also estimate our success in partitioning the galaxy
catalog into group and field subsamples.  We find that our
group-finder correctly identifies 70\% of true group galaxies, with
46\% of the identified group galaxies being interlopers from the
field, while 4\% of identified field galaxies are actually members of
groups (these numbers are 66\%, 43\% and 6\%, respectively, in the
EGS).  A more complete listing of group-finding success and failure
statistics, including different definitions of the completeness and
purity, is given in Table~\ref{tab:catalog_properties}.

To optimize the group-finding parameters, we made use of mock catalogs
constructed using three different N-body simulations, with varying
values of the cosmological parameters $\Omega_M$ and $\sigma_8$.  The
purity and completeness values we estimate from these catalogs vary
quite strongly with the assumed cosmology.  This underscores the
importance of opimizing the group finder on mocks that are consistent
with as broad a range of data as possible, including both cosmological
information and the properties of the galaxy population.  We have
taken care in this work to report statistics from mock catalogs that
match the galaxy data and are consistent with all current cosmological
constraints.  We note that earlier high-redshift group-finding efforts
(\eg, G05, K09, C10) have been optimized on mocks with now-disfavored
cosmologies, so their reported purity and completeness statistics
should be treated with caution.

As we did in G05, we find that the optimized VDM group-finding
algorithm can reconstruct the distribution of groups in velocity
dispersion and redshift for the main DEEP2 sample, above a threshold
in dispersion of $\sigmav \sim 300$\kms, to an accuracy better than
the sample variance in this distribution.  Moreover, our optimal parameter set
can achieve this level of accuracy for all three mock cosmologies
considered here, in spite of the different purity and completeness
statistics obtained for the different mocks.    As shown in
\cite{NMCD02},  the velocity function of groups can be used as a probe
of cosmological paramters; the fact that VDM can accurately
reconstruct this distribution for a range of cosmologies makes the
DEEP2 group catalog presented here a promising dataset for this test,
although a more detailed accounting for sources of
incompleteness and contamination in the group catalog will be
important in any such analysis.

\acknowledgments

We thank C. Marinoni for making the VDM group-finding algorithm
available for our use.  This work was supported in part by NSF grants
AST-0507428, AST-0507483, AST-0071048, AST-0071198, AST-0808133 and
AST-0806732. BFG was supported by the U.S. Department of Energy under
contract number DE-AC02-76SF00515. MCC received support from NASA
through Hubble Fellowship grant \#HF-51269.01-A awarded by the Space
Telescope Science Institute, which is operated by the Association of
Universities for Research in Astronomy, Inc., for NASA, under contract
NAS 5-26555; and from the Southern California Center for Galaxy
Evolution, a multi-campus research program funded by the University of
California Office of Research. The data presented herein were obtained
at the W.M. Keck Observatory, which is operated as a scientific
partnership among the California Institute of Technology, the
University of California and the National Aeronautics and Space
Administration. The Observatory was made possible by the generous
financial support of the W.M. Keck Foundation.  The DEIMOS
spectrograph was funded by a grant from CARA (Keck Observatory), an
NSF Facilities and Infrastructure grant (AST92-2540), the Center for
Particle Astrophysics and by gifts from Sun Microsystems and the
Quantum Corporation.  The DEEP2 Redshift Survey has been made possible
through the dedicated efforts of the DEIMOS staff at UC Santa Cruz who
built the instrument and the Keck Observatory staff who have supported
it on the telescope.  Finally, the authors wish to recognize and
acknowledge the very significant cultural role and reverence that the
summit of Mauna Kea has always had within the indigenous Hawaiian
community.  We are most fortunate to have the opportunity to conduct
observations from this mountain.

\begin{deluxetable}{ccccccc}
\tabletypesize{\small}
\tablewidth{0pt}
\tablecaption{The DEEP2 VDM group catalog (A sample is shown here.
  The full table is available in electronic form in the
    electronic edition of the Journal or from the DEEP2 DR4 web page.)}
\tablehead{\colhead{ID\tablenotemark{a}}&\colhead{R.A.\tablenotemark{b,c}}&\colhead{Decl.\tablenotemark{b,c}}&\colhead{$z$\tablenotemark{b}}&N & \colhead{$\sigma_v$\tablenotemark{d}}& \colhead{H.-P.\tablenotemark{e}}}
\startdata
$ 2052$ & $  02\,  31\, 22.39$ & $+  00\,  35\, 19.6$ &   0.92 &    6 &        878 &   Yes \\
$ 2450$ & $  02\,  26\, 59.36$ & $+  00\,  47\, 31.0$ &   1.01 &    3 &        873 &   Yes \\
$ 2328$ & $  02\,  30\, 33.57$ & $+  00\,  30\, 34.2$ &   1.04 &    3 &        873 &   Yes \\
$ 2151$ & $  02\,  30\, 26.08$ & $+  00\,  38\, 04.5$ &   0.87 &    4 &        819 &   Yes \\
$ 2147$ & $  02\,  27\, 59.14$ & $+  00\,  38\, 47.4$ &   0.93 &    5 &        813 &    No \\
$ 1912$ & $  23\,  30\, 36.04$ & $+  00\,  04\, 05.2$ &   0.96 &    3 &        930 &    No \\
$ 1835$ & $  23\,  32\, 26.54$ & $+  00\,  08\, 05.0$ &   1.21 &    3 &        894 &   Yes \\
$ 1764$ & $  23\,  29\, 41.23$ & $+  00\,  15\, 02.8$ &   0.83 &    3 &        712 &    No \\
$ 2018$ & $  23\,  25\, 50.32$ & $+  00\,  20\, 03.0$ &   1.00 &    2 &        708 &    No \\
$ 1908$ & $  23\,  33\, 15.36$ & $-  00\,  02\, 34.4$ &   0.89 &    2 &        685 &   Yes \\
$ 1416$ & $  16\,  49\, 52.53$ & $+  35\,  05\, 53.6$ &   1.12 &    3 &        960 &   Yes \\
$ 1172$ & $  16\,  51\, 15.36$ & $+  34\,  53\, 19.0$ &   0.82 &    3 &        813 &   Yes \\
$ 1175$ & $  16\,  53\, 09.44$ & $+  35\,  00\, 17.5$ &   0.86 &    4 &        808 &   Yes \\
$ 1186$ & $  16\,  48\, 08.06$ & $+  34\,  56\, 08.6$ &   0.77 &    3 &        807 &   Yes \\
$ 1391$ & $  16\,  52\, 50.72$ & $+  34\,  52\, 24.2$ &   1.27 &    3 &        667 &   Yes \\
$  998$ & $  14\,  18\, 29.23$ & $+  52\,  50\, 40.8$ &   0.73 &    3 &        902 &    No \\
$ 1078$ & $  14\,  15\, 20.47$ & $+  52\,  19\, 29.4$ &   0.91 &    4 &        889 &    No \\
$  334$ & $  14\,  22\, 02.89$ & $+  53\,  25\, 06.2$ &   0.26 &    3 &        853 &   Yes \\
$  857$ & $  14\,  23\, 40.21$ & $+  53\,  31\, 59.3$ &   0.58 &    3 &        847 &   Yes \\
$  142$ & $  14\,  15\, 30.90$ & $+  52\,  16\, 37.4$ &   0.46 &    5 &        845 &   Yes \\
\enddata
\tablenotetext{a}{Unique identification number for each group.}
\tablenotetext{b}{Units of right ascension are hours, minutes, and seconds, and units of declination are degrees, arcminutes, and arcseconds.  All values are given for J2000.0 Equinox.  The Extended Groth Strip is at 14 hours right ascension.}
\tablenotetext{c}{Median value of all galaxies in the group.}
\tablenotetext{d}{Given in km s$^{-1}$.}
\tablenotetext{e}{Groups detected with both the standard and the high-purity parameters are indicated as High-Purity.}
 
\label{tab:groupcatalog}
\end{deluxetable}


\begin{deluxetable}{cc}
\tabletypesize{\small}
\tablewidth{0pt}
\tablecaption{Galaxies in the DEEP2 VDM group catalog and the ID
  numbers of their host groups (A sample is shown here.
  The full table is available in electronic form in  the
    electronic edition of the Journal or from the DEEP2 DR4 web page.)}
\tablehead{\colhead{galaxy ID\tablenotemark{a}} & \colhead{group
ID\tablenotemark{b}}}
\startdata
$ 43024982$ & $ 2052$ \\
$ 43024870$ & $ 2052$ \\
$ 43024803$ & $ 2052$ \\
$ 42027302$ & $ 2052$ \\
$ 42027156$ & $ 2052$ \\
$ 42027036$ & $ 2052$ \\
$ 42021576$ & $ 2053$ \\ 
$ 42021537$ & $ 2053$ \\ 
$ 42015342$ & $ 2054$ \\ 
$ 42015248$ & $ 2054$ \\ 
$ 42014311$ & $ 2054$ \\ 
$ 32010356$ & $ 1562$ \\ 
$ 32010351$ & $ 1562$ \\ 
$ 32010253$ & $ 1562$ \\ 
$ 32010001$ & $ 1562$ \\ 
$ 32019550$ & $ 1564$ \\ 
$ 32019541$ & $ 1564$ \\ 
$ 32019539$ & $ 1564$ \\ 
$ 32019392$ & $ 1564$ \\ 
$ 32100597$ & $ 1565$ \\ 
$ 32019079$ & $ 1565$ \\ 
$ 32019015$ & $ 1565$ \\ 
$ 32018964$ & $ 1565$ \\ 
$ 32018960$ & $ 1565$ \\ 
$ 32018881$ & $ 1565$ \\ 
$ 32018851$ & $ 1565$ \\ 
$ 32018836$ & $ 1565$ \\ 
$ 32018720$ & $ 1565$ \\ 
$ 22026881$ & $ 1166$ \\ 
$ 22019554$ & $ 1166$ \\ 
$ 22019550$ & $ 1166$ \\ 
$ 22027206$ & $ 1167$ \\ 
$ 22026729$ & $ 1167$ \\ 
$ 21028032$ & $ 1168$ \\ 
$ 21027914$ & $ 1168$ \\ 
$ 21027687$ & $ 1168$ \\ 
$ 13102037$ & $ 1010$ \\ 
$ 13057569$ & $ 1010$ \\ 
$ 13057335$ & $ 1010$ \\ 
$ 12025635$ & $ 1011$ \\ 
$ 12025616$ & $ 1011$ \\ 
$ 11019701$ & $ 1012$ \\ 
$ 11019555$ & $ 1012$ \\ 
\enddata
\tablenotetext{a}{Unique galaxy identification number in the DEEP2 catalog.}
\tablenotetext{b}{Unique identification number for each group, as in Table~\ref{tab:groupcatalog}.}
 
\label{tab:galcatalog}
\end{deluxetable}

\bibliographystyle{apj}
\bibliography{}

\appendix
\section{Analytical technique for predicting the velocity function of groups}

One of the primary stated purposes of this study
is to create a catalog of massive systems that can be used to
constrain cosmological parameters by measuring the observed
distribution of groups in redshift and velocity dispersion.  As a
service to future users of the catalog, we outline here a method for
computing this distribution as a function of cosmological parameters.
All of the calculations outlined below can be straightforwardly
performed to include redshift dependence; we do not explicitly include
this for succinctness.

The goal is to constrain the mass function of dark-matter halos, whose
dependence on cosmology is well understood theoretically, using the
measured velocity dispersion of groups as a proxy for halo mass.
Doing this requires an analytical technique for converting from the
halo mass function to the velocity function, \ie, the distribution of
galaxy groups as a function of their observed velocity dispersion
$\sigmav$.  When forecasting the constraints that might be possible
from DEEP2, \citet{NMCD02} developed a method for converting from the
mass function to the \emph{halo} velocity function, as would be
measured using the dark matter velocity dispersion $\sigdm$, but that
analysis did not account for the substantial additional scatter in the
$M-\sigmav$ relation arising from observational selection effects and
the limited number of galaxies used to calculate each group's
$\sigmav$ (as shown in Figure~\ref{fig:dilution}).

In any case, the first step is to compute the halo mass function
$dn/dM$ and convert it to the halo velocity function
$dn/d\sigdm$.  The theoretical halo mass function has a relatively
simple dependence on cosmological parameters and has been calibrated
against  N-body simulations to an accuracy better than required for
DEEP2 comparisons \citep[e.g.,][and references
therein]{Tinker08}.  The relation between halo mass $M$ and
$\sigdm$ is also well calibrated from N-body simulations
\citep{Evrard07}, so we can use this relation to convert from $dn/dM$
to $dn/d\sigdm$ directly.\footnote{In principle there is
  some scatter about the $M$--$\sigdm$ relation, so the
  conversion should actually take the form of a convolution of the
  mass function with the distribution of $\sigdm$ at fixed $M$,
  but in practice this scatter is much smaller than the
  observationally induced scatter we discuss below, so it can be
  neglected in favor of a simple change of variables.}  Then the
observed velocity function takes the form of a convolution to account
for observational scatter:
\begin{equation}
dn/d\siggal = \int_0^\infty dn/d\sigdm
P(\siggal|\sigdm) d\sigdm,
\label{eqn:velfn_convol}
\end{equation}
where $\siggal$ is the velocity dispersion measured using
the observed galaxies in the group.  

The remaining task is then to compute the conditional probability
distribution function (PDF) in Equation~\ref{eqn:velfn_convol}.  Provided that the
number of galaxies $N$ per halo is known, and assuming that these
galaxies randomly sample the dark matter velocity field (i.e., that
there is no velocity bias) and that the underlying velocity
distribution is normal, the problem reduces to computing the
distribution of measured dispersions obtained by sampling a Gaussian
$N$ times.    The form of this distribution depends on the estimator
we use to compute the measured dispersion.  For the usual standard deviation
estimator, it can be computed analytically \citep{Kenney51}:
\begin{equation}
P_\mathrm{sd}(\siggal|\sigdm, N) \propto
\frac{1}{\sigdm}\left(\frac{\siggal}{\sigdm}\right)^{N-2}  
\exp{\left[-\frac{N}{2}\left(\frac{\siggal}{\sigdm}\right)^2\right]},
\label{eqn:prob_stddev}
\end{equation}
with the normalization being set by
the condition $\int_0^\infty P\,d\siggal=1$.

Our preferred estimator, however, is the gapper,
Equation~\ref{eqn:gapper}.  The sorting of the data points required by
that estimator makes analytical computation of the desired conditional
PDF intractable for $N>2$.  However, for the $N=2$ case, we can derive
a closed form, via a calculation very similar to the one that produced
Equation~\ref{eqn:prob_stddev}:
\begin{equation}
P_G(\siggal|\sigdm, N=2) \propto
\frac{1}{\sigdm} 
\exp{\left[-\frac{1}{\pi}\left(\frac{\siggal}{\sigdm}\right)^2\right]}.
\label{eqn:prob_gap_2}
\end{equation}
The notable similarity between the $N=2$ cases of $P_G$
and $P_\mathrm{sd}$ suggests a possible approximate form for
$P_G$ in the case of general $N$,
\begin{equation}
P_G(\siggal|\sigdm, N) \propto
\frac{1}{\sigdm}\left(\frac{\siggal}{\sigdm}\right)^{N-2}  
\exp{\left[-\frac{\alpha(N)}{2}\left(\frac{\siggal}{\sigdm}\right)^2\right]}, 
\label{eqn:prob_gap_fit}
\end{equation}
where $\alpha(N)$ is a free parameter that needs to be determined by
numerical calulations.
Monte-Carlo experiments indicate that the one-parameter fitting
formula,  
Equation~\ref{eqn:prob_gap_fit}, is a remarkably good model for the
$\siggal$ PDF.  By fitting this 
function to Monte-Carlo simulated distributions over the range $2\le
N\le 25$, we find that the parameter $\alpha(N)$ is well described by 
\begin{equation}
\alpha(N) = N-2\left(1-\frac{1}{\pi}\right)
\left(1+\frac{1-e^{-N/2+1}}{10}\right). 
\label{eqn:alpha_fit}
\end{equation}

Armed with the above formula for $P_G(\siggal|\sigdm, N)$, we can
write down a form for the desired conditional PDF:
\begin{equation}
P(\siggal|\sigdm) = \sum_{\Nobs=2}^\infty [P(\Nobs|\sigdm) P(\siggal|\sigdm, \Nobs)],
\label{eqn:probsig_overall}
\end{equation}
 where
$\Nobs$ is the number of \emph{observed} galaxies in a particular
halo.  The lower limit of the sum obtains because systems with
$\Nobs<2$ do not qualify as groups;  we note that this sum
thus also implicitly encodes the halo selection function in full
detail, provided that we have an accurate form for $P(\Nobs|\sigdm)$.

What remains, then, is to determine $P(\Nobs|\sigdm)$.  This can be
done in principle using sufficiently realistic mock catalogs to
measure the desired PDF directly.  Since this is likely to be noisy,
however, it may be preferable to use the techniques of the halo
model: one can simply measure the observed HOD
$\bar{N}_\mathrm{obs}(M)$ and assume a simple well-motivated model for the
scatter about this mean, such as the model of \citet{Zheng05},
who use Poisson scatter for satellites and a nearest-integer
distribution for centrals.  We note that, since this is the HOD of
\emph{observed} galaxies, including all spectroscopic selection
effects, the form of the HOD will be different from the one that
is usually assumed---for example, the mean number of centrals will
always be less than unity---but the scatter model should
still be applicable.

Thus, given sufficiently realistic mock catalogs, it is possible to
predict the observed group velocity function analytically.  The
question of what constitutes sufficient realism in the mocks warrants
brief discussion, though.  Since we are using the mocks to compute the
observed HOD, we would like them to accurately \emph{(a)} reflect the
true \emph{total} HOD $\bar{N}(M)$ in the Universe and \emph{(b)}
include all selection effects relevant to the dataset being used.  The
first requirement is equivalent to demanding that the mocks reproduce
the observed two-point correlation function of galaxies, and the
second necessitates careful modeling of galaxy properties to match the
data.  As discussed above and in Gerke et al. (in prep.), the mock
catalogs we use here meet the latter requirement but not the former.
However, as also discussed in Gerke et al. (in prep.), the
improvements to the mock catalogs that are likely to improve the
clustering agreement are also likely only to impact the low-mass
portion of the HOD, where $\bar{N}\la 1$.  Since the above formalism
only relies on objects with $\Nobs>2$ , there will be little
contribution from this inaccurate portion of the HOD.  Thus, it is
possible that the mocks we used in this study are sufficiently
accurate to carry out the modeling described here; however, this
assumption would need to be tested in some detail before robust
cosmological constraints could be derived from the velocity function
of DEEP2 groups.  

\begin{figure}
\centering
\epsfig{width=3.5in, file=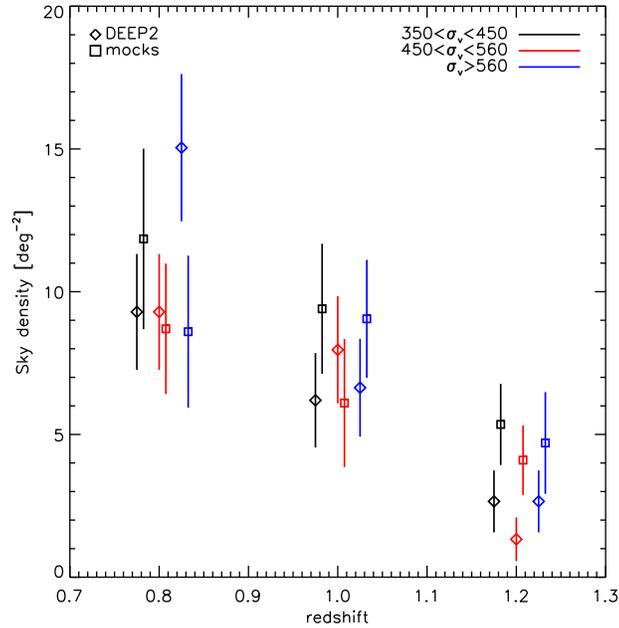}
\caption{On-sky density of VDM galaxy groups in the DEEP2 survey and
  in the mocks, using the same redshift and velocity-dispersion bins
  that we used in Figure~\ref{fig:velfn_resid}.  DEEP2 points show the
  density of groups in the three high-redshift DEEP2 fields, and error
  bars show Poisson uncertainties on these measurements.  Mock data
  points show the mean counts from 20 one-square-degree mock
  lightcones.  Mock error bars show the estimated DEEP2 sample
  variance, which is computed as the standard deviation of the values
  from the 20 mocks reduced by a factor of 0.65 to account for the
  three independent DEEP2 fields, two of which are only partially
  observed. }
\label{fig:append_compare}
\end{figure}

A simple way to test this within the confines of the present study is
to compare the abundance of DEEP2 groups in bins of $\sigmav$ to the
abundance of groups detected by VDM in the mocks.  If we assume that
the mocks are a sufficiently accurate representation of the galaxy
population in groups, then this test constitutes a consistency check
between DEEP2 and the background $\Lambda$CDM cosmology of the Bolshoi
mocks (which is very close to the current best fit to the available
data).  Figure~\ref{fig:append_compare} shows this comparison.  Data
points show the on-sky density of groups detected with VDM in DEEP2
and in the mocks, for groups with $\sigmav \ge 350$ \kms, using the
same binning in redshift and $\sigmav$ that we did in
Figure~\ref{fig:velfn_resid}.  The DEEP2 points are measured from the
three high-redshift DEEP2 fields, and the error bars are Poisson
uncertainties.  The mock data points and error bars show the mean and
standard deviation in the counts from 20 different mock lightcones,
with the error bars reduced by a factor of 0.65 to account for the
reduction in sample variance expected by combining the three independent
high-redshift DEEP2 fields, given that two of them were only partially
observed. We computed this factor using the sample-variance estimation
techniques developed in \citet{ND00}.  

The mock and DEEP2 group abundances are in good statistical agreement
at $z\la 1$, suggesting that DEEP2 is in good agreement with the
Bolshoi background cosmology and mock galaxy modeling in this regime.  In
the highest redshift bin, all three DEEP2 measurements lie 
below the mocks at roughly the $2\sigma$ level.  It is important to
note that the three DEEP2 data points at each redshift are strongly
correlated with one another, since the entire halo mass function will
be scaled up or down in overdense or underdense regions of the
universe.  Hence, the similar disagreement for all three high-redshift
$\sigmav$ bins does not necessarily indicate a systematic
difference.  We can conceive of three possible explanations for this
discrepancy, which we list in decreasing order of plausibility:
\emph{(1)} the mock catalogs may not capture the DEEP2
galaxy population at $z\ga 1$ sufficiently accurately to model
the group selection function, \emph{(2)} the DEEP2 survey may probe
a volume of the universe that is underdense at the $\sim 2\sigma$
level at $z\ga 1$, or \emph{(3)} the growth of cosmic structure in the
Bolshoi mocks is inconsistent with the real universe, indicating a
problem with the current concordance cosmology.  Construction of a
DEEP2 mock catalog that more accurately reproduces the measured
two-point function would help to settle this question.

\end{document}